\documentclass{jfm}

\usepackage{graphicx}
\usepackage{natbib}
\usepackage{amsmath}
\usepackage{amssymb}
\usepackage{bm}
\usepackage[usenames,dvipsnames]{xcolor}

\title[Resonant and near-resonant interactions in rotating
    turbulence]{Quantifying resonant and near-resonant interactions in
    rotating turbulence}
\author[P. Clark di Leoni and P. D. Mininni]
{Patricio Clark di Leoni
    \thanks{Email address for correspondence: clark@df.uba.ar}
    and Pablo D. Mininni}

\affiliation{Departamento de F\'\i sica, Facultad de Ciencias 
    Exactas y Naturales, Universidad de Buenos Aires and IFIBA, CONICET, 
    Ciudad Universitaria, 1428 Buenos Aires, Argentina.}

\pubyear{2010}
\volume{650}
\pagerange{119--126}
\date{?; revised ?; accepted ?. - To be entered by editorial office}
\begin{document}

\maketitle

\begin{abstract}
    Nonlinear triadic interactions are at the heart of our understanding
    of turbulence. In flows where waves are present modes must not only
    be in a triad to interact, but their frequencies must also satisfy
    an extra condition: the interactions that dominate the energy
    transfer are expected to be resonant. We derive equations that
    allow direct measurement of the actual degree of resonance of each
    triad in a turbulent flow. We then apply the method to the case of
    rotating turbulence, where eddies coexist with inertial waves. We
    show that for a range of wave numbers, resonant and near-resonant
    triads are dominant, the latter allowing a transfer of net energy
    towards two-dimensional modes that would be inaccessible
    otherwise. The results are in good agreement with approximations
    often done in theories of rotating turbulence, and with the
    mechanism of parametric instability proposed to explain the
    development of anisotropy in such flows. We also observe that, at
    least for the moderate Rossby numbers studied here, marginally
    near-resonant and non-resonant triads play a non-negligible role
    in the coupling of modes.
\end{abstract}

\section{Introduction} 

Understanding the nature of nonlinear interactions is at the core of the
problem of turbulence. It has been known for quite some time that the
transfer of energy between scales in a turbulent flow involves groups
of three spatial modes, namely a {\it triad} $({\bf k}, {\bf p}, {\bf
    q})$ such that ${\bf k} = {\bf p} + {\bf q}$. The concept was first
introduced by \cite{Kraichnan58}, where he showed that triadic
interactions are conservative, and later used this representation to
formulate his Direct Interaction Approximation theory. Later,
\cite{Lee75} analysed triads according to geometric arguments. The
subsequent growth of computing power allowed analysis of triadic
interactions in direct numerical simulations, and \cite{Domaradzki90a}
showed that while energy transfer in the inertial range is mostly
local in wave number space, nonlocal triads (i.e., triads involving
modes with disparate wave numbers) can have large amplitudes (although
they are less numerous, and thus the flux is dominated by the local
triads, see \citet{Eyink09,Aluie09}). An important result was obtained
by \cite{Waleffe92}, who decomposed the velocity field in terms of
helical modes, and also analysed locality aspects of the nonlinear
interactions. This allowed him to identify which triads, when
isolated, contribute to a direct energy transfer (energy going from
large to small scales), and which to an inverse transfer (energy going
from small to large scales).

The concept of triadic interactions remains relevant to present
date. Nonlinear interactions were analysed by 
\cite{Mininni06,Mininni08}, by studying local and nonlocal triads, and
the shell-to-shell energy transfer functions in wavenumber
space. Recently, \cite{Cheung14} formulated an exact representation of 
the nonlinear triads using a combination matrix, and were able use it
along with minimal assumptions to obtain the Kolmogorov spectrum from
the Navier-Stokes equation. The helical decomposition of
\cite{Waleffe92} was also used recently to build ``decimated''
versions of the Navier-Stokes equation \citep{Biferale13}, where the
nonlinear terms in the equation are split into contributions from
each kind of triad, and which can then be turned on or off to see how
they affect the energy cascade. In a similar way, \cite{Moffatt14}
analysed the effect of triad truncation on the velocity and vorticity
field of the Euler equation. Finally, decimation models in which
  wave-wave-wave interactions and wave-vortex-wave interactions were
  differentiated have been used to study rotating stratified
  turbulence \citep{Remmel10,Hernandez-duenas14}.

In flows with restitutive forces and in which waves can be present
(e.g., in rotating and/or stratified flows, or in magnetohydrodynamics),
an important concept arises which is that of resonants triads. These are
triads $({\bf k},{\bf p},{\bf q})$ which also satisfy the resonant
condition $\omega ({\bf k}) = \omega ({\bf p}) +\omega ({\bf q})$, with
$\omega ({\bf k})$ being the dispersion relation of the waves. If a flow
is dominated by rapidly varying waves, non-resonant interactions should,
in principle, die out in front of resonant ones, thus leaving the bulk
of the nonlinear energy transfer to the resonant triads. This has been
exploited in theories of weak turbulence (i.e., in systems in which the
flow is completely given by a superposition of interacting dispersive
waves), as done for interfacial waves in fluids or for waves in plasmas
\citep{Zakharov,Nazarenko} with varying degrees of success
\citep{Newell11}. Experimental evidence of such resonant wave
interactions has been found, e.g., in capillary wave turbulence
\citep{Aubourg15,Aubourg16} and in gravity-capillary waves
\citep{Haudin16}.

The Coriolis force in rotating flows gives rise to inertial waves which
in experiments and in simulations coexist with eddies
\citep{Staplehurst08,Bokhoven08}.  As a result, although weak turbulence
theories can give some insight into the energy transfer mechanisms
\citep{Galtier03,Nazarenko11}, more general formulations of wave
turbulence in the strong regime are needed to describe the flow
\citep{Cambon89,Cambon97}. The first attempts to study resonant triads
in these flows were carried out by \cite{Newell69}, who studied how
these triads become the preferred energy transfer mechanism and its
implication for the formation of planetary zonal flows. Extensions of
Rapid Distortion Theory and of the Eddy-Damped Quasi-Normal Markovian
closure to rotating turbulent flows rely heavily on resonant
interactions, and can correctly capture the development of anisotropy in
rotating turbulence \citep{Cambon89,Cambon97,Bellet06}. Moreover, this
approach is useful to understand how the flow becomes quasi-two
dimensional, with energy in three-dimensional modes being transferred
preferentially towards modes with smaller vertical wavenumber through a
subset of the resonant triads. Within the framework of the helical
decomposition, \cite{Waleffe93} also considered the resonant triads, and
pointed out that a parametric instability may be the mechanism behind
the preferential transfer of energy towards quasi-two dimensional modes:
the resonant condition $\omega ({\bf k}) = \omega({\bf p}) + \omega
({\bf q})$ is more easily satisfied by modes with small vertical
wavenumber, thus being preferred by the nonlinear coupling. This
tendency in the energy transfer towards quasi-two dimensionalisation has
been confirmed both in numerical simulations \citep{Sen12,Horne13} and
in laboratory experiments \citep{Campagne15}. However, the parametric
instability mechanism of \citet{Waleffe93} is valid for isolated
triads; in a real turbulent flow, in which each triad is coupled to a
miriad of other triads, it is unclear whether this is the actual
mechanism responsible for the quasi-two dimensionalisation.

Moreover, wave turbulence theories are valid when the wave period is
much shorter than the eddy turnover time. As a result, many of these
arguments fail when the Rossby number is moderate, or when the
vertical wavenumber approaches zero (as there are no waves in these
modes), and thus they cannot predict whether energy is transferred
into pure two-dimensional modes. Also, wave turbulence theories are
inhomogeneous in scale space, and even for small Rossby numbers there
can exist a sufficiently small scale such that the time scales of the
eddies and of the waves become of the same order thus violating its
hypothesis \citep{Pouquet10,Mininni12}. In recent years, several
efforts were made to detect inertial waves in rotating turbulence,
quantify their energy, and identify their role in the anisotropic
transfer of energy. Some relied on the fact that bounded domains have
resonant frequencies which can be  spotted in a temporal spectrum
\citep{Bewley07,Rieutord12,Lamriben11}. Others analysed temporal
decorrelation functions to determine at which scales wave action was
predominant \citep{Favier10,Clark14}. \cite{Campagne15} identified the
presence of inertial waves by analysing the two-point spatial
correlation of the time transformed velocity fields obtained from PIV
measurements. Also, the space and time resolved energy spectrum
was calculated both numerically \citep{Clark14,Clark15b} and
experimentally \citep{Yarom14,Campagne15}. While all this evidence
points to a strong presence of waves in rotating flows, and thus of
resonant interactions, studies of the contribution of resonant and
near-resonant triads in rotating turbulence are scarce.

Recently, experimental evidence of three-wave resonant interactions has
been found in a rotating flow \citep{Bordes12}. In numerical
simulations, \cite{Chen05} compared rotating flows computed in grids of
$128^3$ spatial points with simulations of the Navier-Stokes equation in
two dimensions, and concluded that resonant triads play a more dominant
role as rotation is increased, while they also raised concerns on the
validity of wave turbulence arguments for the long time dynamics of the
flow. Following \cite{Waleffe92}, \cite{Smith05} considered numerical
simulations of truncated systems in which only some interactions were
preserved, to identify which triads were responsible for the development
of anisotropy. The authors concluded that near-resonant interactions
were needed to reproduce the quasi-two dimensionalisation of the flow,
while non-resonant triads reduce this anisotropic transfer. More
recently, \cite{Alexakis15} analysed a large numerical dataset of
rotating flows and concluded that the dynamics of the quasi-two
dimensional component of the flow can only be correctly captured if
near-resonant and non-resonant interactions are taken into account.
Also recently, \cite{Gallet15} showed that two-dimensional flows are
preferred solutions of rotating flows for small enough Rossby number,
indicating that a description of the energy transfer solely in terms
of resonant triads has limitations even in the limit of very strong
rotation. The role of near-resonant interactions is also
  important to understand the limit of infinite domains, see 
  \citet{Cambon04,Chen05,Bourouiba08} for discussions. However,
direct measurements of resonant interactions in turbulent flows are
hard to find, owing primarily to the massive amount of data that needs
to be extracted and analysed from either experiments or numerical
simulations.

The aim of this paper is to directly quantify how different triads
contribute to the energy transfer in rotating turbulence.  It is worth
mentioning that a similar analysis was performed recently on
experimental data of gravity-capillary waves measured on the surface of
a liquid \citep{Aubourg16}, where interactions are also between
three waves. The analysis, based on phenomenological arguments,
allowed direct identification of resonant interactions. Here we
develop a theoretical formalism for three-waves interactions in a
rotating flow that allows explicit derivation of third order
correlation functions between modes. We do this by deriving a 
{\it contribution function}, a function that measures the 
contribution of each triad to the total energy transfer as a function
of  the wavenumber and frequency, and a {\it normalised contribution
  function} that measures the characteristic timescale at which an
interaction takes place. Both allow the measurement of how relevant
and how well tuned (i.e., how resonant) a given triad is. We then use
these tools to analyse results from direct numerical simulations of
rotating turbulence. The formalism can be extended to other systems
with three or more wave interactions.

We start in Sec.~\ref{theoback} with a brief explanation of the nature
of nonlinear interactions in the Navier-Stokes equations, followed by
a description of our numerical simulations. Then, in
Sec.~\ref{explicacion} we derive the aforementioned contribution
function, which we then use to analyse the data from numerical
simulations of rotating turbulence in Sec.~\ref{numresults}. Finally,
in Sec.~\ref{conclusions} we present our conclusions.

\section{Resonant triads} 
\label{theoback}

\subsection{Nonlinear interactions in Navier-Stokes}

In a rotating frame, the Navier-Stokes equations for an incompressible
fluid with velocity ${\bf u}$ and under the action of a mechanical
forcing ${\bf F}$ read
\begin{gather}
    \frac{\partial {\bf u}}{\partial t} = - ({\bf u} \cdot {\bm \nabla})
    {\bf u}
         - 2 \Omega \hat{z} \times {\bf u} 
        - {\bm \nabla} {\cal P} + \nu \nabla^2 {\bf u} + {\bf F} ,
    \label{momentum}
    \\
    {\bm \nabla} \cdot {\bf u} =0 ,
    \label{incompressible}
\end{gather}
where ${\cal P}$ is the total pressure (including the centrifugal
force, and normalised by the uniform fluid mass density), $\hat{z}$ is
parallel to the rotation axis, $\Omega$ is the rotation frequency, and
$\nu$ is the kinematic viscosity. The Reynolds number, defined as
usual as $\textrm{Re} = UL/\nu$ (where $U$ is the r.m.s.~velocity and
$L$ is the energy injection scale) quantifies the strength of the
nonlinear term against viscous damping.

Using the incompressibility condition given by
Eq.~(\ref{incompressible}), and for ${\bf F}=0$, the Fourier transform
of Eq.~\eqref{momentum} can be written as
\begin{equation}
    \label{transformed}
    \left( \frac{\partial}{\partial t} 
        - \nu k^2 + 2 \Omega \mathbb{P}_{\bf k} \hat{z} \times \right) {\bf u_k} 
    = -i \mathbb{P}_{\bf k} \sum_{{\bf p}+{\bf q}={\bf k}} ({\bf u_p}
        \cdot {\bf q}) {\bf u_q} ,
\end{equation}
where $[\mathbb{P}_{\bf k}]_{ij} = \delta_{ij} - k_i k_j /k^2$
    is the projector operator, which projects in the direction
    perpendicular to ${\bf k}$ to enforce incompressibility. All
terms on the l.h.s.~of the equation are linear in ${\bf u_k}$ and do
not couple modes with different ${\bf k}$. So, after transforming
the nonlinear term in Eq.~\eqref{momentum}, the resulting
convolution on the r.h.s.~of this equation tells us that only modes
${\bf p}$ and ${\bf q}$ in triads satisfying 
${\bf k}={\bf q} + {\bf p}$ can give or receive energy from the mode
with wave vector ${\bf k}$. Note this is not a unique property 
of the Navier-Stokes equation but of any nonlinear equation with
quadratic nonlinearities.

\subsection{Rotating flows and relevant time scales} 

In the presence of rotation (or of other restitutive forces), waves
can be excited with well defined frequencies for each wave vector, 
given by the dispersion relation of the waves $\omega({\bf k})$. For a
rotating flow, the dispersion relation of inertial waves is 
\begin{equation}
    \omega({\bf k}) = \pm \frac{2 \Omega k_z}{k}.
    \label{dispersion}
\end{equation}
We can then define the Rossby number as
\begin{equation}
    \textrm{Ro} = \frac{U}{2L \Omega} ,
\end{equation}
which measures the ratio of the rotation period to the turnover time
of the large-scale eddies. As a result, for small Rossby number we can
expect waves to be faster than eddies, at least for a range of
scales. Indeed, in the absence of forcing and viscous effects, 
Eqs.~\eqref{momentum} and \eqref{incompressible} have waves with
dispersion relation \eqref{dispersion} as exact solutions.

We can thus define several relevant time scales, as in a turbulent
flow one can identify different characteristic times for each
possible interaction. The sweeping of the small scale eddies
(of size $\sim 1/k$) by the large scale flow is described by the
{\it sweeping time} 
\citep{Chen89}
\begin{equation}
    \tau_S \sim \frac{1}{U k} .
    \label{sweeping}
\end{equation}
Note that sweeping does not result in a transfer of energy
  across scales. Sweepping corresponds to the advection of the small
  scale eddies by a large scale flow, and can act even in the absence
  of a mean flow (i.e., just a random evolution of the modes at large
  scales can result in {\it random sweeping}). The advection of the
  small scale eddies in real space corresponds to a rotation of the
  Fourier modes by $e^{iUkt}$. The interaction of similar sized
  eddies, which result in nonlinear transfer of energy, is described
by the {\it nonlinear time scale},
\begin{equation}
    \tau_{NL} \sim \frac{1}{k\sqrt{k E(k)}} ,
    \label{nonlintime}
\end{equation}
where $E(k)$ is the energy spectrum of the flow. Finally, the time
scale for the interaction of waves modes is expected to be proportional
to the wave period, i.e.,
\begin{equation}
    \tau_\omega \sim \frac{1}{|\omega({\bf k})|} .
    \label{wavetime}
\end{equation}
While in homogenous and isotropic turbulence the dominant Eulerian
time is the sweeping time \citep{Chen89}, when waves are present the
dominant time scale can be either the sweeping time, the nonlinear
time, or the wave period depending on which is the fastest at a given
scale \citep{Favier10,Servidio11,Clark14,Clark15b}. As a result, these
time scales imply that depending on the shape of the energy
spectrum $E(k)$, the approximation that waves are faster than the
eddies for small enough $\textrm{Ro}$ may break down for sufficiently 
large wave numbers, or for a subset of the Fourier modes (e.g., for
modes with small vertical wavenumber). Below we present a more
detailed estimation of which is the dominant time scale for each mode
in our  numerical simulations.

\subsection{Resonant interactions} 

For modes for which the waves are much faster than the eddies, we can
assume wave dynamics dominate the evolution, while the eddies
contribute to a slow modulation of the amplitude of the waves. Thus,
we can write ${\bf u_k} = {\bf U_k} e^{i \omega_{\bf k} t}$. In
practice, this approximation is done after decomposing the modes 
${\bf u_k}$ into the helical eigenstates ${\bf h}_\pm ({\bf k})$ of
the linearised Eq.~\eqref{momentum}, 
${\bf u_k} = a_+({\bf k}) {\bf h}_+ ({\bf k}) + 
    a_-({\bf k}) {\bf h}_- ({\bf k})$,
where the subindices $+$ and $-$ correspond to the two possible
polarisation of the waves, see e.g., \cite{Waleffe93}. However, for
the purpose of the following discussion it is better to work in terms
of ${\bf u_k}$, as those modes are more easily accessed in numerical 
simulations.

Replacing in Eq.~\eqref{transformed} we obtain
\begin{equation}
    \left( \frac{\partial}{\partial t} - \nu k^2 + 
    2 \Omega \mathbb{P}_{\bf k} \hat{z} \times \right) {\bf U_k}
  = -i \mathbb{P}_{\bf k} \sum_{{\bf p}+{\bf q}={\bf k}} ({\bf U_p}
    \cdot {\bf q}) {\bf U_q} 
    e^{-i [\omega (\bf k) - \omega ({\bf p}) - \omega ({\bf q})]t}.
\end{equation}
Integrating over several periods of the waves, the nonlinear term can
give a non-negligible energy transfer only if triads are resonant,
i.e., if $\omega ({\bf k}) = \omega ({\bf p}) + \omega ({\bf q})$. In
practice, near-resonant triads with 
\begin{equation}
\gamma_r ({\bf k}, {\bf p}, {\bf q}) = 
    \frac{
    \min\{|\omega ({\bf k}) \pm \omega ({\bf  p}) \pm \omega 
        ({\bf q})|\}}
    {2 \Omega} =
    \cal{O}(\textrm{Ro}) ,
\end{equation}
are also expected to be relevant \citep[see, e.g.,][]{Alexakis15}. We
will call $\gamma_r$ the resonance factor, as it measures how close to
resonance a given triad is in the framework of wave turbulence
theory. The minimum and the plus-minus signs added in the latter
equation are due to the fact that our modes ${\bf u_k}$ mix both
polarisations of the inertial waves.

\section{Contribution of nonlinear triads to the energy transfer and
  to the eddy decorrelation} 
\label{explicacion}

In the traditional picture of turbulence, energy is transferred
towards smaller scales as the eddy gets sufficiently deformed (and
thus, decorrelated in time) by the interaction with other eddies. As
we are interested in understanding the role of the waves in the energy
transfer, we need an expression for the contribution of each triad to
the decorrelation of individual modes (and thus, to the distribution
of energy per wavenumber). To do this we define 
${\bf u_k} = {\bf u_k}(t)$ and ${\bf u'}_{\bf k} = {\bf u_k} (t')$
with $t'=t-\tau$, and multiply Eq.~\eqref{transformed} by 
${\bf u'}^*_{\bf k}$. After averaging over the time $t'$ and assuming
the system is in a turbulent steady state, we obtain
\begin{equation}
    \left( \frac{\partial}{\partial t} - \nu k^2 \right)
        \left< {\bf u'}^*_{\bf k} \cdot {\bf u_k} \right>_{t'} 
        + 2 \Omega \left< {\bf u'}^*_{\bf k} \cdot
          ( \hat{z} \times {\bf u_k} ) \right>_{t'} =
        -i \sum_{{\bf p}+{\bf q}={\bf k}} \left< {\bf u'}^*_{\bf k}
        \cdot ({\bf u_p} \cdot {\bf q}) {\bf u_q} \right>_{t'} ,
    \label{firststep}
\end{equation}
where the projector $\mathbb{P}_{\bf k}$ was dropped as the dot
product with ${\bf u'}^*_{\bf k}$ ensures only components of the
terms perpendicular to ${\bf k}$ survive (as 
${\bf k} \cdot {\bf u'}^*_{\bf k} = 0 $ from the incompressibility
condition). In Eq.~(\ref{firststep}) we also assume that the complex
conjugate is added (this must be assumed in all the following
equations).

The viscous term on the l.h.s.~of Eq.~(\ref{firststep}) is just
responsible for damping of the correlations in a viscous time scale
which grows as the Reynolds number. Thus, for large Reynolds numbers
its effect can be neglected in comparison to the wave and the
nonlinear time scales. After assuming the system is in a turbulent
steady state, we can then rewrite this equation in terms of
functions that depend only on the time lag $\tau$ as
\begin{equation}
    \frac{\partial}{\partial \tau} \Gamma_{\bf k}(\tau) 
    + 2\Omega \Delta_{\bf k}(\tau) =
    -i \sum_{{\bf p}+{\bf q}={\bf k}} \Theta({\bf k},{\bf q},{\bf p},\tau)
    \label{eqgamt}
\end{equation}
where
\begin{equation}
    \Gamma_{\bf k} (\tau) = \langle {\bf u}^*_{\bf k}(t') \cdot {\bf
        u_k}(t'+\tau) \rangle_{t'}
\end{equation}
and
\begin{eqnarray}
    \Delta_{\bf k} (\tau) &=& \langle {\bf u}^*_{\bf k}(t') \cdot
    \left[ \hat{z} \times {\bf u_k}(t'+\tau) \right] \rangle_{t'} \\
{} &=&
    \langle u_y^*({\bf k}, t') u_x({\bf k},t'+\tau) \rangle_{t'}
    - \langle u_x^*({\bf k},t') u_y({\bf k},t'+\tau) \rangle_{t'}
\end{eqnarray}
are time correlation functions for the mode ${\bf k}$ (with $\Gamma$
the usual time correlation function used in isotropic and homogeneous
turbulence),  and the third-order time correlation is
\begin{equation}
    \Theta ({\bf k},{\bf p},{\bf q},\tau) = \langle {\bf u}^*_{\bf k} (t') \cdot 
        \left[ {\bf u_p}(t'+\tau) \cdot {\bf q} \right] {\bf u_q}(t'+\tau)
        \rangle_{t'} .
        \label{thetadef}
\end{equation}
The function $\Delta_{\bf k}$ is zero for $\tau = 0$, and can be
removed from these equations for all time lags if correlation
functions are written for the amplitudes of the helical eigenstates
${\bf h_\pm}({\bf k})$. Moreover, even in terms of the Fourier modes
of the velocity ${\bf u}_{\bf k}$, after adding the complex conjugate
and assuming the system is in a turbulent steady state (i.e., that the
statistical properties of the signals are homogeneous in time), this
function can be neglected.

In a turbulent flow, the correlation function $\Gamma_{\bf k}$ is thus
expected to decrease to $1/e$ of its value at $\tau=0$ on a 
timescale that may be either $\tau_S$, $\tau_{NL}$, or
$\tau_\omega$. This decorrelation results from the interaction with
all triads, with the contribution from each triad measured by the
triple correlation $\Theta ({\bf k},{\bf q},{\bf p},\tau)$. Thus,
computation of this function should allow identification of the
dominant interactions responsible for the energy cascade discriminated
by time scale. Note also that for $\tau =0$, $\Theta$ reduces to the
usual transfer function $T({\bf k},{\bf p},{\bf q})$ that measures the
strength of each individual triad 
\citep{Kraichnan58,Domaradzki90a,Waleffe92,Mininni11b}.

As the Fourier transform of the correlation function is proportional
to the power spectrum, we have
\begin{equation}
\widehat{\Gamma_{\bf k} (\tau)} = 2 E({\bf k},\omega) ,
\end{equation}
and as from the property of derivatives of Fourier transformed
functions
\begin{equation}
\widehat{\frac{\partial}{\partial \tau} \Gamma_{\bf k}(\tau)} =
    -2 i \omega E({\bf k},\omega) ,
\end{equation}
we thus arrive to
\begin{equation}
\label{ekwtheta}
2 \omega E({\bf k},\omega) = \sum_{{\bf p}+{\bf q}={\bf k}}
    \widehat{\Theta} ({\bf k},{\bf q},{\bf p},\omega).
\end{equation}
Note that $\widehat{\Theta}$ quantifies how much each triad 
$({\bf k}, {\bf p}, {\bf q})$ and each frequency $\omega$ contribute
to the space and time (four-dimensional) energy spectrum. Also, how
well tuned $\widehat{\Theta}$ is around a given $\omega({\bf k})$
can be used to identify how close to resonance a triad actually is. We
will thus call $\widehat{\Theta}$ the  {\it contribution
  function}.

We can gain further insight into the meaning of $\widehat{\Theta}$ by
studying the case of a fluid in which only waves are present. In this
particular case, we can write 
${\bf u}_{\bf k} = {\bf U}_{\bf k} e^{i \omega_{\bf k} t}$, and we can
neglect any slow dependence of ${\bf U}_{\bf k}$ in time. Bearing
aside normalisation factors for simplicity, we have
\begin{eqnarray}
\widehat{\Theta}  ({\bf k},{\bf q},{\bf p},\omega) 
    &=& \int_{-\infty}^{\infty} e^{i \omega \tau}
               \left< {\bf u}^*_{\bf k} (t') \cdot \left[ {\bf
               u_p}(t'+\tau) \cdot {\bf q} \right] {\bf
               u_q}(t'+\tau) \right>_{t'} \mathrm{d} \tau
               \nonumber \\
    &=& \int_{-\infty}^{\infty} e^{i (\omega +\omega_{\bf p} +
               \omega_{\bf q}) \tau} \left< {\bf U}^*_{\bf k}
               \cdot \left( {\bf U_p} \cdot {\bf q}\right) {\bf
               U_q} \, e^{-i (\omega_{\bf k} -\omega_{\bf p} -
               \omega_{\bf q}) t'}\right>_{t'} \mathrm{d} \tau
               \nonumber \\ 
    &=& {\bf U}^*_{\bf k} \cdot ({\bf U_p} \cdot {\bf q}) {\bf U_q} 
               \, \delta(\omega-\omega_{\bf k}) .
\label{eq:delta}
\end{eqnarray}
So in this case $\widehat{\Theta}$ only contributes to the frequency
$\omega = \omega({\bf k})$, and thus only resonant triads contribute
to $E({\bf k},\omega)$. In practice $\widehat{\Theta}$ will not always
be sharply peaked around $\omega({\bf k})$, as shown below. The width
of the peak can therefore be used to quantify how resonant a triad is.

It is much easier, both conceptually and practically, to work with a
symmetrised $\widehat{\Theta}$, namely
\begin{equation}
    \widehat{\Theta}^S ({\bf k},{\bf q},\omega) = \frac{1}{2} \left[
\widehat{\Theta} ({\bf k},{\bf q},{\bf p}={\bf k}-{\bf q},\omega) +
\widehat{\Theta} ({\bf k},{\bf p}={\bf k}-{\bf q},{\bf q},\omega)
\right].
\end{equation}
From here on after every mention of $\Theta$ and its Fourier transform
will be in this symmetrised form. The superscript $S$ shall therefore
be dropped.

\section{Numerical results} 
\label{numresults}

\begin{figure}
    \centering
    \includegraphics[scale=0.5]{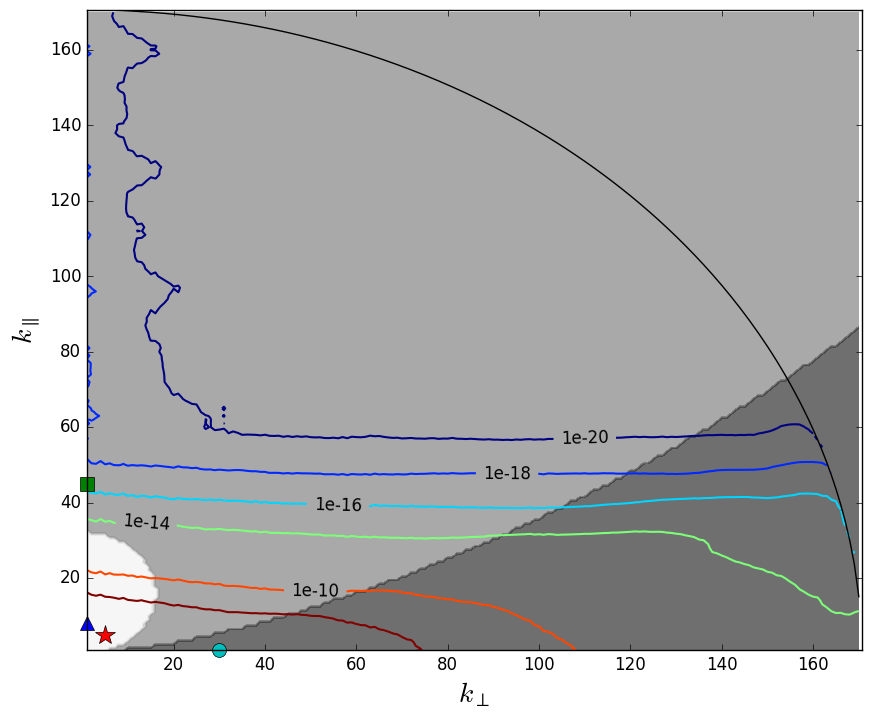}
    \caption{Contour levels of the energy spectrum as a function of
      parallel and perpendicular wave numbers, in the simulation of
      rotating turbulence with $\Omega=8$. Note the anisotropy of the
      spectrum, with most of the energy accumulating near modes with
      $k_\parallel \approx 0$. The white region corresponds to modes
      with $\tau_\omega<\tau_s<\tau_{NL}$ (i.e., modes dominated by
      the waves), the light grey region to modes with
      $\tau_s<\tau_\omega<\tau_{NL}$, and the dark grey region to
      modes with $\tau_s<\tau_{NL}<\tau_{\omega}$ (see text for
      details). Grey regions thus correspond to modes for which
      sweeping gives the fastest time scale (with dark grey indicating
      ``slow'' modes). The location of four modes relevant for the
      analysis are marked in the figure: the ``wave'' modes 
      ${\bf k}=(0,0,8)$ (marked with a blue triangle) and 
      ${\bf k}=(0,5,5)$ (marked with a red star), a ``slow''
      (two-dimensional) mode in the dark grey region 
      ${\bf k}=(0,30,0)$ (marked with a cyan circle), and the mode 
      ${\bf k}=(0,0,45)$ in the light grey region (marked with a green
      square).}
    \label{ekk}
\end{figure}

\subsection{Numerical simulations} 

The code GHOST \citep{Gomez05,Mininni11} is used to solve 
Eqs.~\eqref{momentum} and \eqref{incompressible} using  a parallel
pseudo-spectral method with a second order Runge-Kutta scheme for the
time evolution. The $2/3$-rule is used for dealiasing. As will be seen
below, computation of the contribution function requires high cadence
I/O in time, and a significant amount of storage (note spatial
information needs to be saved with twice the frequency of the fastest
waves in the system). As a result, only simulations with moderate
resolution can be performed. Here we present two simulations using
grids of $N^3=512^3$ points in a three-dimensional periodic box.

Both simulations are identical except for the value of $\Omega$. In
one of the simulations $\Omega=4$, while in the other $\Omega=8$. The
simulations were started from the fluid at rest, and energy was
injected via a mechanical forcing. We chose a Taylor-Green forcing of
the form
\begin{eqnarray}
    {\bf F} &=& F_0 \left[ \sin(k_{\textrm{TG},x} x)
                \cos(k_{\textrm{TG},y} y) 
                \cos(k_{\textrm{TG},x} z) \hat{x} \right.
                \nonumber \\
         {} && \left. -\cos(k_{\textrm{TG},x} x)
                \sin(k_{\textrm{TG},y} y) 
                \cos(k_{\textrm{TG},z} z) \hat{y} \right] ,
\end{eqnarray}
with $F_0 = 0.277$, ${\bf k}_\textrm{TG} = (1,1,1)$ (which results in
$L=2\pi/k_\textrm{TG}=2\pi/\sqrt{3}$), and $\nu=6.5\times10^{-4}$ in
dimensionless units (for unit velocity and a box of length $2\pi$).
While other forcings will presumably produce similar results,
    Taylor-Green forcing was chosen because it has been reported to
    result in a larger amplitude of wave modes when compared with
    random-in-time isotropic forcing \citep{Clark14}. The system was
let to reach a turbulent steady state with $U\approx0.9$, which
translates to a Reynolds number of approximately 5000, and a Rossby
number of $0.03$ for $\Omega=4$ and of $0.015$ for $\Omega=8$. 
Once in t he turbulent steady state the simulations were allowed to run
for over 12 large scale turn over times, the time span over which the
following analysis was carried on.

\subsection{Energy spectrum and decorrelation times}

\begin{figure}
    \centering
    \includegraphics[scale=0.6]{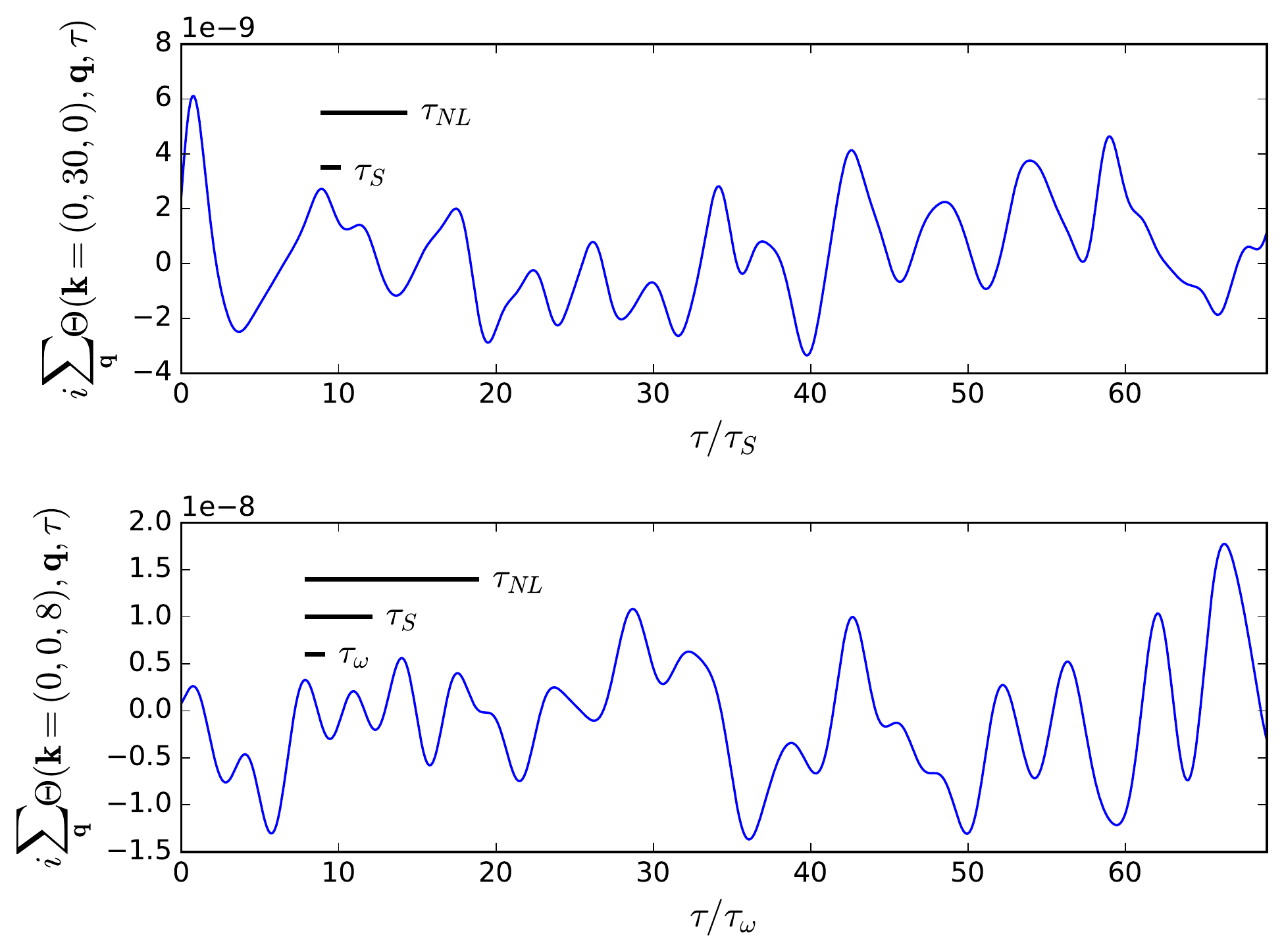}
    \caption{Partial reconstruction of 
        $-{\partial \Gamma_{{\bf k}}}/{\partial \tau}$ via the sum over
        ${\bf q}$ of the third-order correlator 
        $\Theta({\bf k}, {\bf q}, \tau)$, from Eq.~\eqref{eqgamt}
        (computation of the real part by adding the complex conjugate
        is implied). The two panels show, from top to bottom, the
        results for ${\bf k}=(0,30,0)$ and for ${\bf k}=(0,0,8)$. The
        time lag $\tau$ is normalised by the dominant time scale
        ($\tau_s$ in the top panel, and $\tau_\omega$ in the bottom
        panel); the different timescales in these units are also
          shown as a reference by the horizontal bars in each figure.
        We recover the expected behaviour for the correlation
        functions, ${\bf k}=(0,0,8)$ (which is a fast mode) gets
        locked to the wave period, while ${\bf k}=(0,30,0)$ (which is
        slow) evolves in the sweeping time scale. The integral of
        these functions gives the correlation function, which decays 
        rapidly on the dominant time scale.}
    \label{thetat}
\end{figure}

\begin{figure}
    \centering
    \includegraphics[scale=0.6]{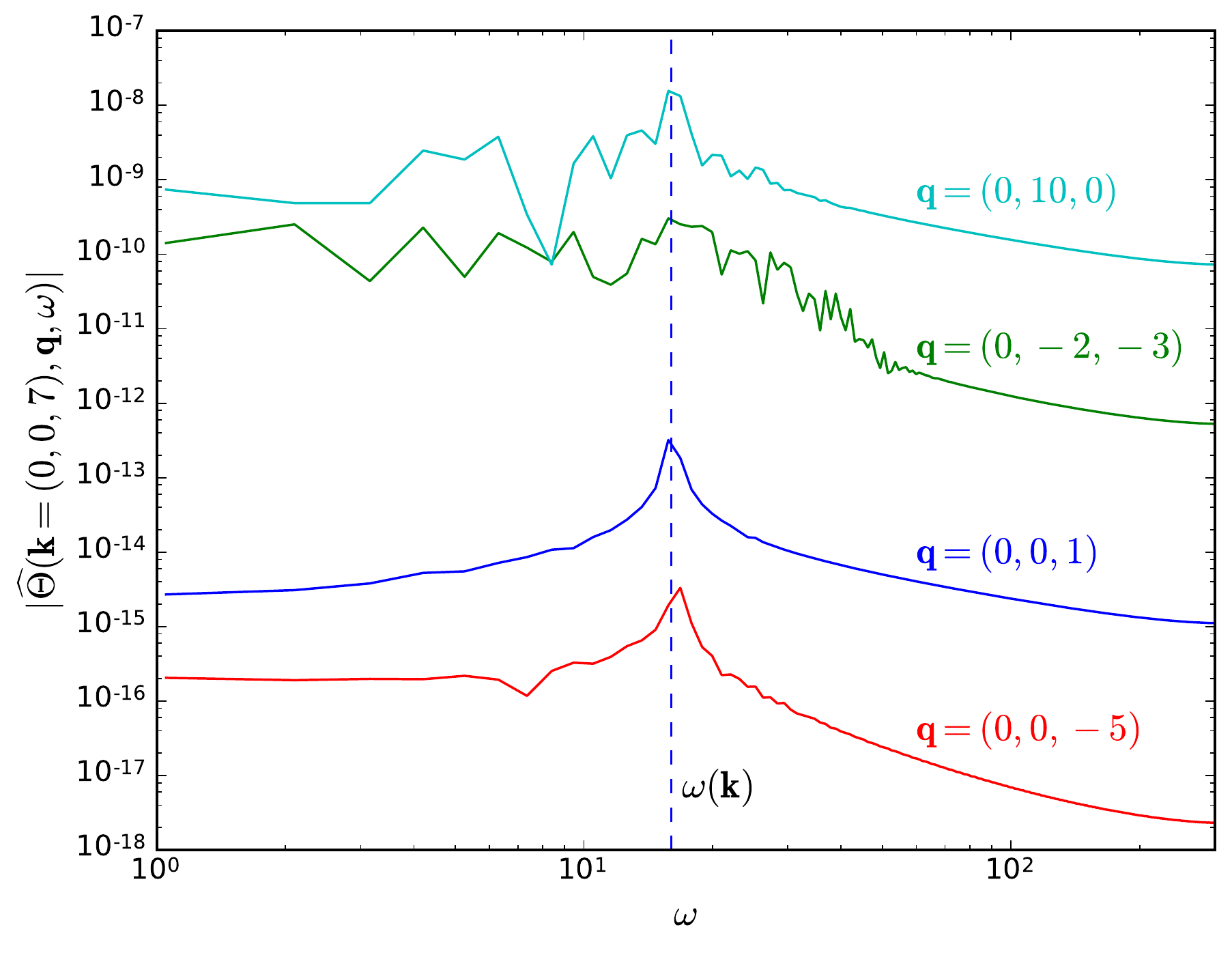}
    \caption{$\vert \widehat{\Theta}({\bf k}=(0,0,8),{\bf
            q},\omega)\vert$ as a function of $\omega$, for a fixed
          fast mode ${\bf k}$, and for different values of ${\bf
            q}$. Interactions with other wave modes (blue and red
          curves) show a well tuned spectrum centred around the wave 
          frequency of the mode ${\bf k}$, indicating interactions are
          close to resonance. On the other hand, interactions with
          eddy modes (green and cyan curves) display a wider
          spectrum. All subsequent analyses of the contribution
          function focus on its maximum amplitude and 
          on how well tuned each triad is (i.e., on the width of the
          peak around the maximum).}
    \label{thetaw}
\end{figure}

Before proceeding to the analysis of the contribution function, we
first discuss some general properties of the simulations. In
Fig.~\ref{ekk} we show contour levels of the axisymmetric energy
spectrum for the simulation with $\Omega=8$, as a 
function of the perpendicular  and parallel wave numbers (with the
perpendicular and parallel directions defined with respect to the axis
of rotation). Whilst in isotropic turbulence one would expect these
contours to be circular, the effect of rotation in these flows impose
a clear anisotropy, with a preferred accumulation of energy in modes
with $k_\parallel \approx 0$ as predicted by
\cite{Cambon89}, \cite{Cambon97}, and \cite{Waleffe92}. Moreover, a
significant fraction of the energy is in modes with $k_\parallel = 0$,
for which resonant interactions cannot account for.

Three different regions are shaded in Fig.~\ref{ekk}. The white region
corresponds to modes with $\tau_\omega<\tau_s<\tau_{NL}$. This is the
region of ``fast'' (or ``wave'') modes, for which the period of the
waves is the fastest time scale. The grey region corresponds to modes
with $\tau_s<\tau_\omega<\tau_{NL}$. Although these modes are often
considered to be ``fast'', as shown in \cite{Clark14} these modes are
decorrelated in a timescale which is of the order of the sweeping
time. In other words, in the Eulerian frame, the dominant time scale
for these modes is given by the sweeping, which is the shortest
available time. Finally, in the dark grey region the modes have
$\tau_s<\tau_{NL}<\tau_{\omega}$. This is the region of ``slow'' 
modes for which the eddies are faster than the waves.
The three shaded regions are shown only as a reference. To
  compute the value of $\tau_{NL}$ at each wave number using 
  Eq.~(\ref{nonlintime}), an estimation of $E(k)$ is needed. For
  simplicity, instead of using the measured spectrum, we use the
  phenomenological expression for non-helical rotating turbulence 
  $E(k)\sim \epsilon^{1/2} \Omega^{1/2} k^{-2}$ 
  \citep{Zhou95,Muller07,Mininni12}. In \citet{Clark14} it was shown,
  from direct computation of the decorrelation times, that this choice
  results in a good estimation of the dominant time scale for modes
  laying in the inertial range (i.e., at small and intermediate wave
  numbers). At large wavenumbers, where the spectrum drops
  exponentially as a result of viscous damping, $\tau_{NL}$ departs
  from this estimation. However, we will not consider modes in the
  viscous range, for which also the viscous damping time can be
  relevant.

The ordering of the time scales described above has implications for
the behaviour of the time correlation function 
$\Gamma_{\bf k}(\tau)$. In the white region of Fig.~\ref{ekk}, 
$\Gamma_{\bf k}(\tau)$ is expected to decay to $1/e$ of its value for
$\tau=0$ in one wave period of the mode with wave vector ${\bf k}$ 
\citep{Favier10,Clark14}. This time scale is what we define as the
decorrelation time: after this time, the mode ${\bf k}$ has
significantly decorrelated from its previous state. As already
mentioned, in the two grey regions the function 
$\Gamma_{\bf k}(\tau)$ decays to $1/e$ of its value for $\tau=0$ in a
time equal to $\tau_s$ \citep{Clark14}. We will thus consider modes in
these three regions to compute the third order correlators 
$\Theta({\bf k}, {\bf p}, {\bf q}, \tau)$ and
$\hat{\Theta}({\bf k}, {\bf p}, {\bf q}, \omega)$. In particular, in
Fig.~\ref{ekk} we indicate two fast modes ${\bf k}=(0,5,5)$ and
$(0,0,8)$, a ``swept'' mode $(0,0,45)$, and a slow mode
$(0,30,0)$. These modes will be used in several examples below.

\begin{figure}
    \centering
    \includegraphics[width=6.5cm]{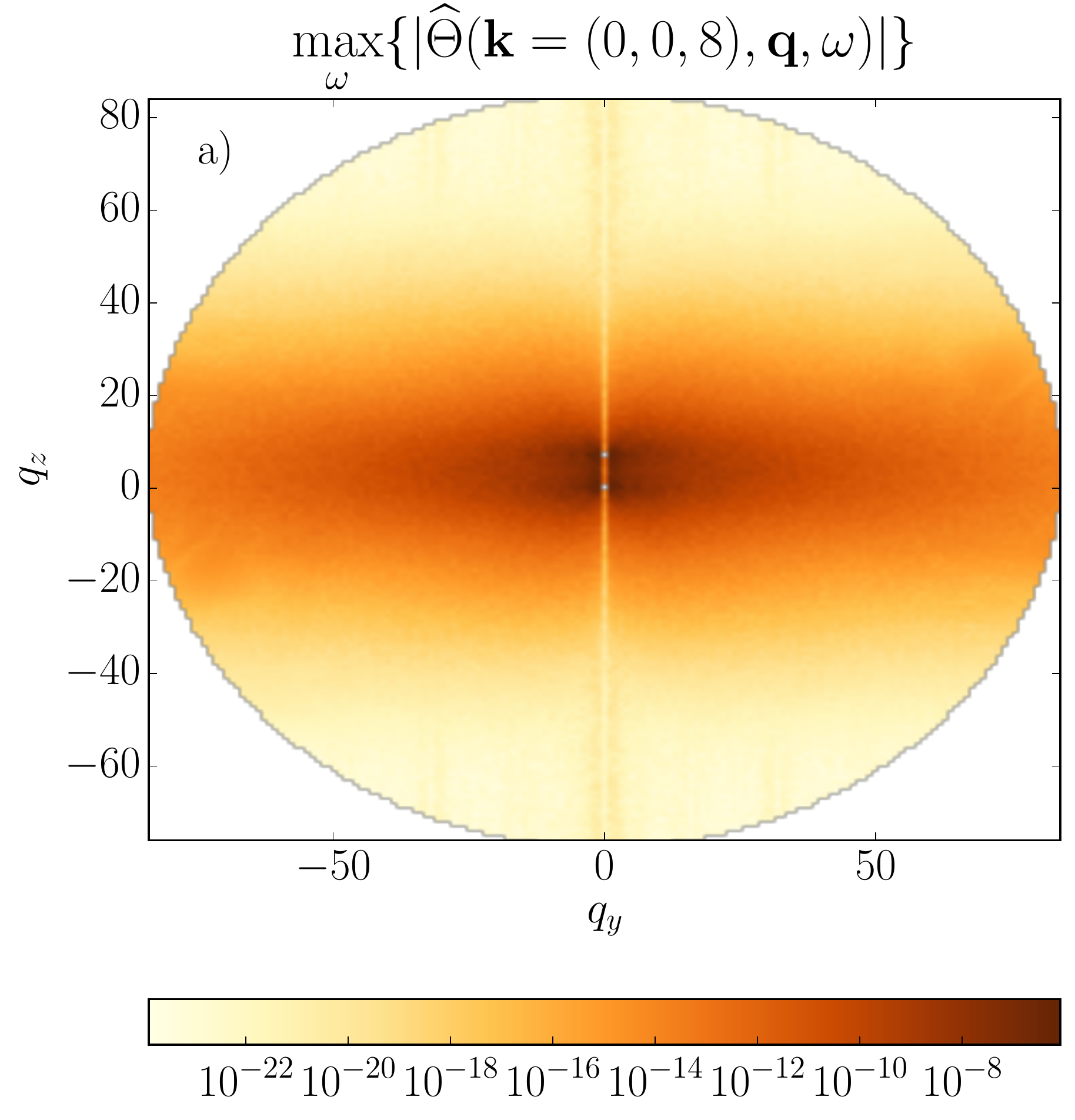}
    \includegraphics[width=6.5cm]{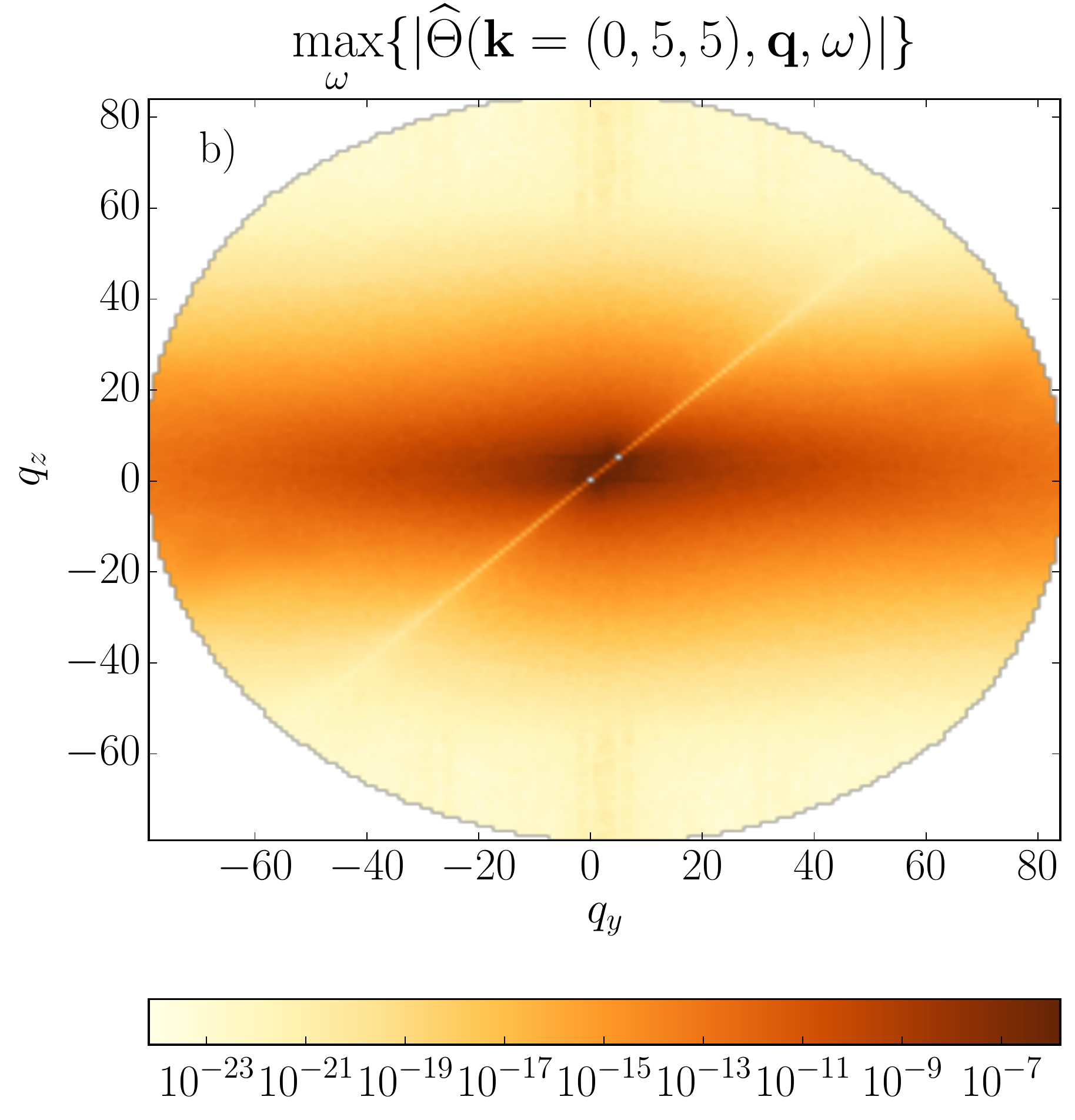}
    \caption{Intensity (as a function of ${\bf q}$) of
      the maximum of the contribution function, 
      $\max_\omega \{\vert \widehat{\Theta} ({\bf k},{\bf q},\omega)
      \vert\}$. In each panel, ${\bf k}$ is fixed, and the maximum of
      $|\widehat{\Theta}|$ is plotted for all available triads by
      varying ${\bf q}$. Two ${\bf k}$ modes are considered, 
      {\it a)} ${\bf k}=(0,0,8)$ and {\it b)} $(0,5,5)$, both
      dominated by waves. The black dots in the centre indicate the
      modes ${\bf q}=\pm{\bf k}$. Two prominent features arise. One is
      the effect of the anisotropy of the flow, as triads with larger
      amplitudes are distributed along horizontal bands (i.e.,
      coupling the ${\bf k}$ modes with modes with smaller
      vertical wave numbers). The other is the defect along the line
      ${\bf q} = \alpha {\bf k}$, as collinear modes do not contribute
      to the triads in an incompressible fluid.}
    \label{peak_contribution}
\end{figure}

\begin{figure}
    \centering
    \includegraphics[width=6.5cm]{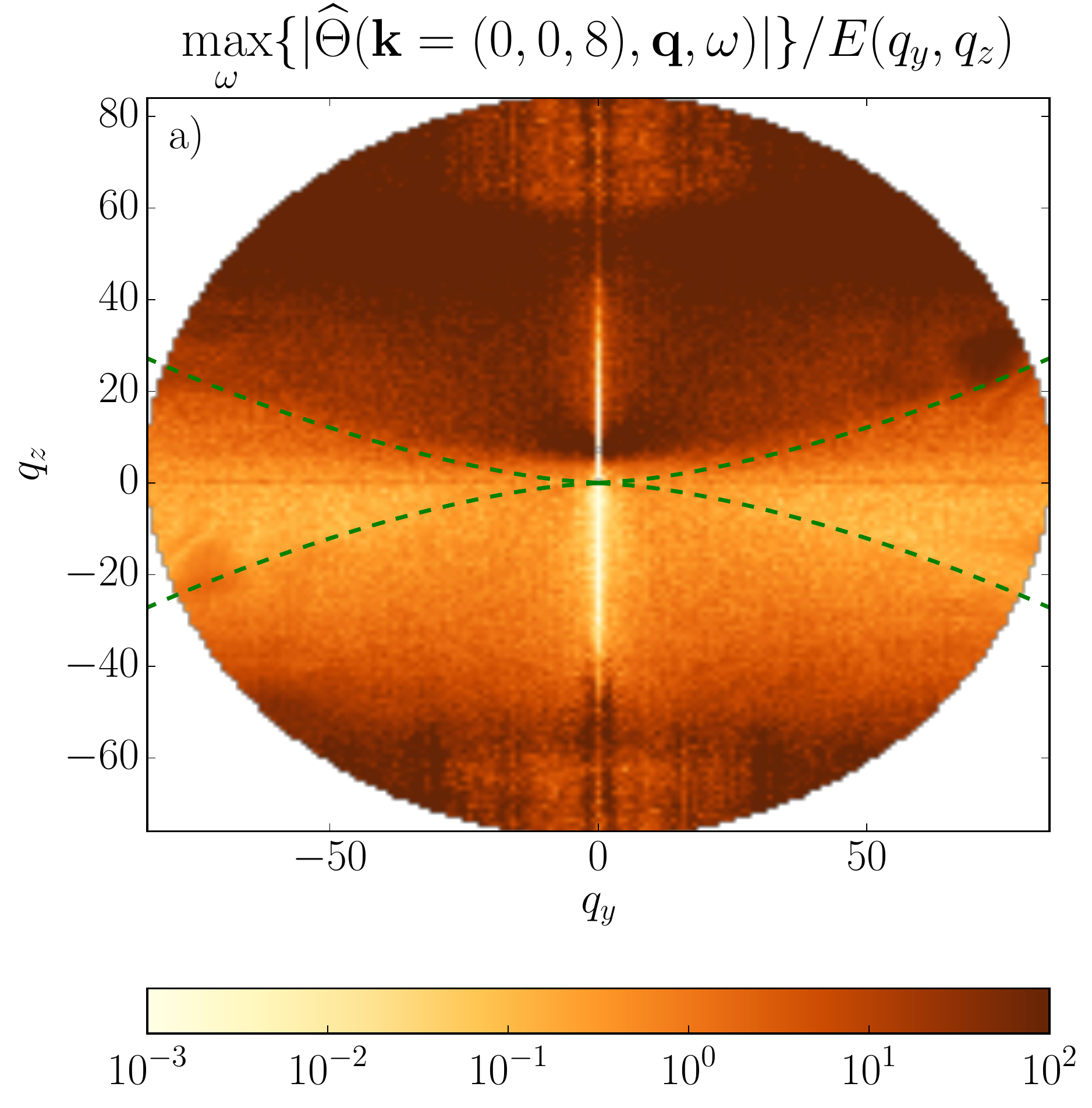}
    \includegraphics[width=6.5cm]{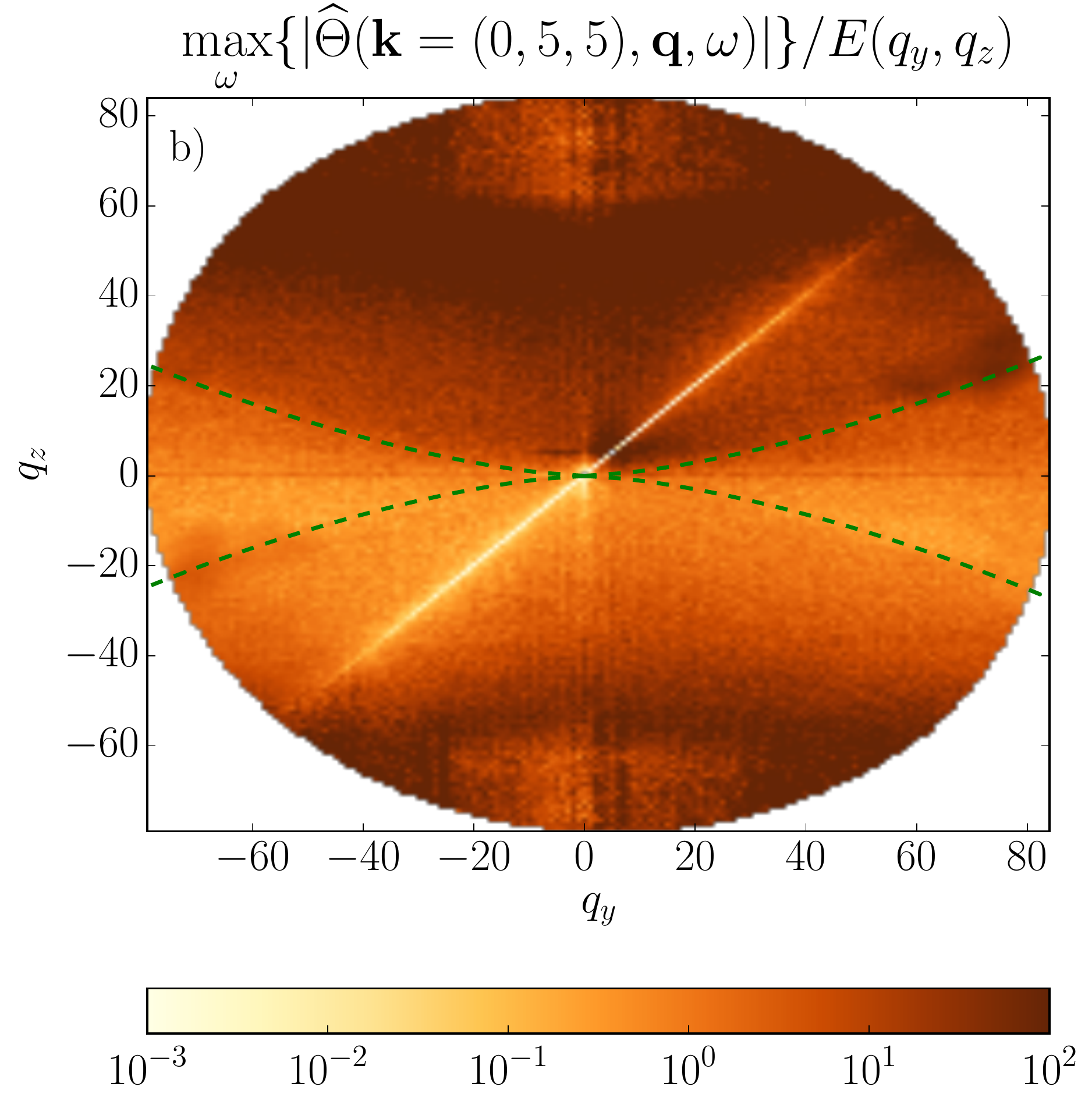}
    \caption{Intensity of the peak values of the
      normalised contribution function for each triad, given by 
      $\max_\omega \{\vert \widehat{\Theta} ({\bf k},{\bf q},\omega)
        \vert\} /E(q_y,q_z)$. In each panel, ${\bf k}$ is
      fixed for two modes dominated by waves: {\it a)} 
      ${\bf k}=(0,0,8)$, and {\it b)} ${\bf k} = (0,5,5)$. The
      dashed lines represent the modes with 
      $\tau_\omega = \tau_{NL}$. Modes with $\tau_\omega<\tau_{NL}$
      (those above the upper dashed line) have higher frequencies (i.e.,
      shorter time scales), and are thus preferred. Note however there
      is a non-negligible leakage towards slow modes ${\bf q}$ with
      $\tau_\omega \gtrsim\tau_{NL}$ (i.e., modes slightly below the
      dashed upper curve).}
\label{peak_frequency}
\end{figure}

\subsection{Analysis of third-order time correlators}

Equation \eqref{eqgamt} indicates that the nonlinear interaction with
all triads gives rise to the time decorrelation of the mode at a given
${\bf k}$. In other words, the apparently random contribution of
all nonlinear triads results in the deformation of the structure to
the point that the mode decorrelates with itself, and thus transfers
its energy to other modes in the allowed triads. As a result,
$\Gamma_{\bf k}(\tau)$ decreases for short increments $\tau$, and then
fluctuates around zero. $-\partial \Gamma_{\bf k}(\tau)/\partial \tau$
should then start from zero for $\tau=0$, increase to a maximum, and
then fluctuate with the dominant time scale of the mode. Figure
\ref{thetat} shows a partial reconstruction of 
$-{\partial \Gamma_{{\bf k}}}/{\partial \tau}$ by computing a partial
sum over ${\bf q}$ of the $\Theta({\bf k}, {\bf q}, \tau)$ function
for ${\bf k}=(0,0,8)$ and for ${\bf k}=(0,30,0)$. The partial
  reconstruction is done using Eqs.~\eqref{eqgamt} and
  \eqref{thetadef}, i.e., we sum $\Theta$ over the subset of ${\bf p}$
  and ${\bf q}$ modes available for the analysis and that satisfy the
  relation ${\bf p} + {\bf q} = {\bf k}$. Due to the large amount of
data required for this computation, we only consider modes in the
$(k_x=0,k_y,k_z)$ plane, and therefore we only sum over the triads
with ${\bf p} \cdot \hat{x} = {\bf q} \cdot \hat{x}=0$, 
resulting in the partial reconstruction mentioned
above. Nonetheless, this suffices to get the expected behaviour for
the time derivative of the decorrelation functions. For the mode 
${\bf k}=(0,0,8)$, which is a fast mode, the time derivative gets
locked to the wave period, while for ${\bf k}=(0,30,0)$, which is a
slow mode, the dominant time scale is the sweeping time. Here and in
the following, except when duly noted, all results shown are for the
$\Omega=8$ simulation.

\begin{figure}
    \centering
    \includegraphics[width=6.5cm]{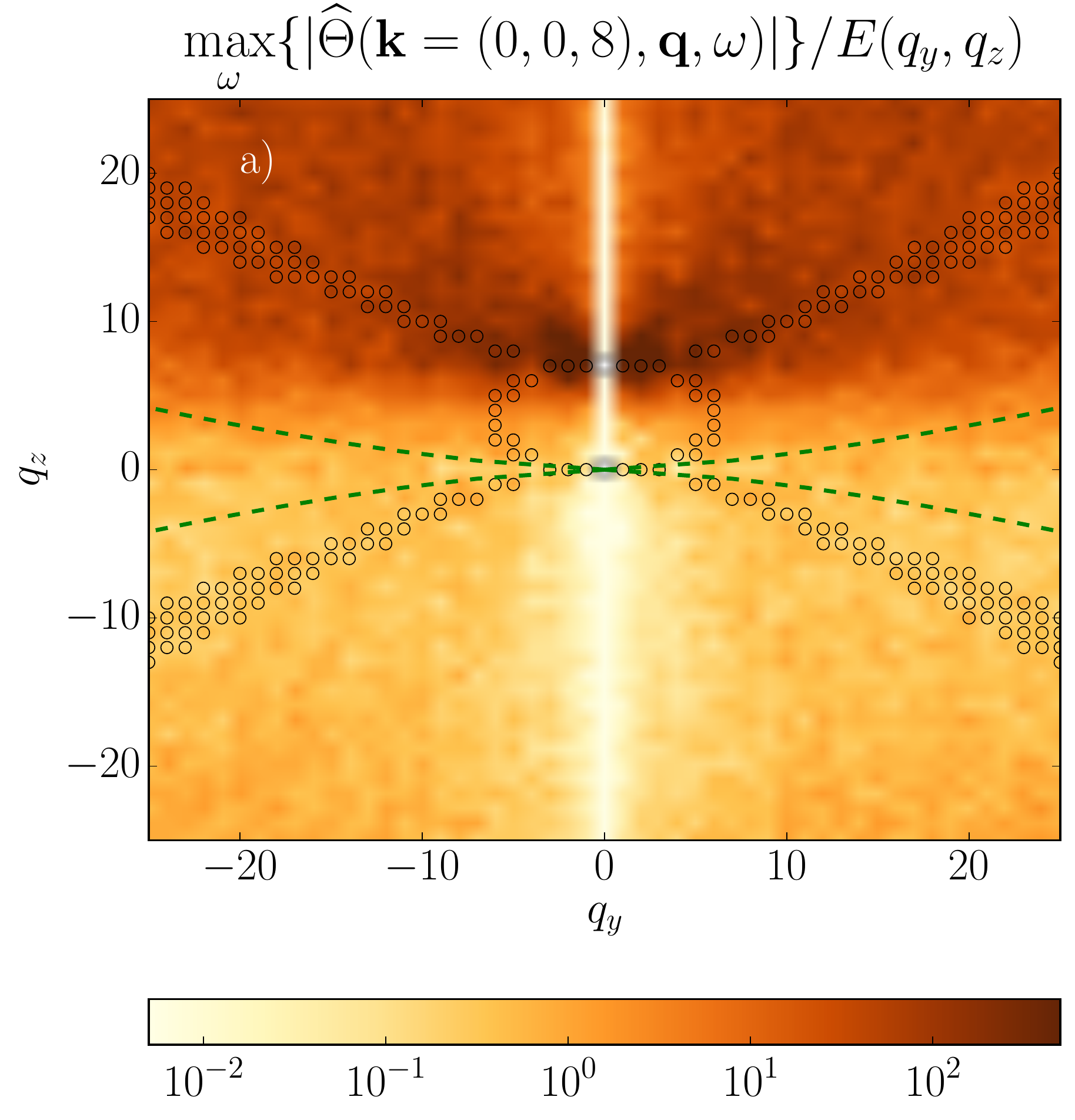}
    \includegraphics[width=6.5cm]{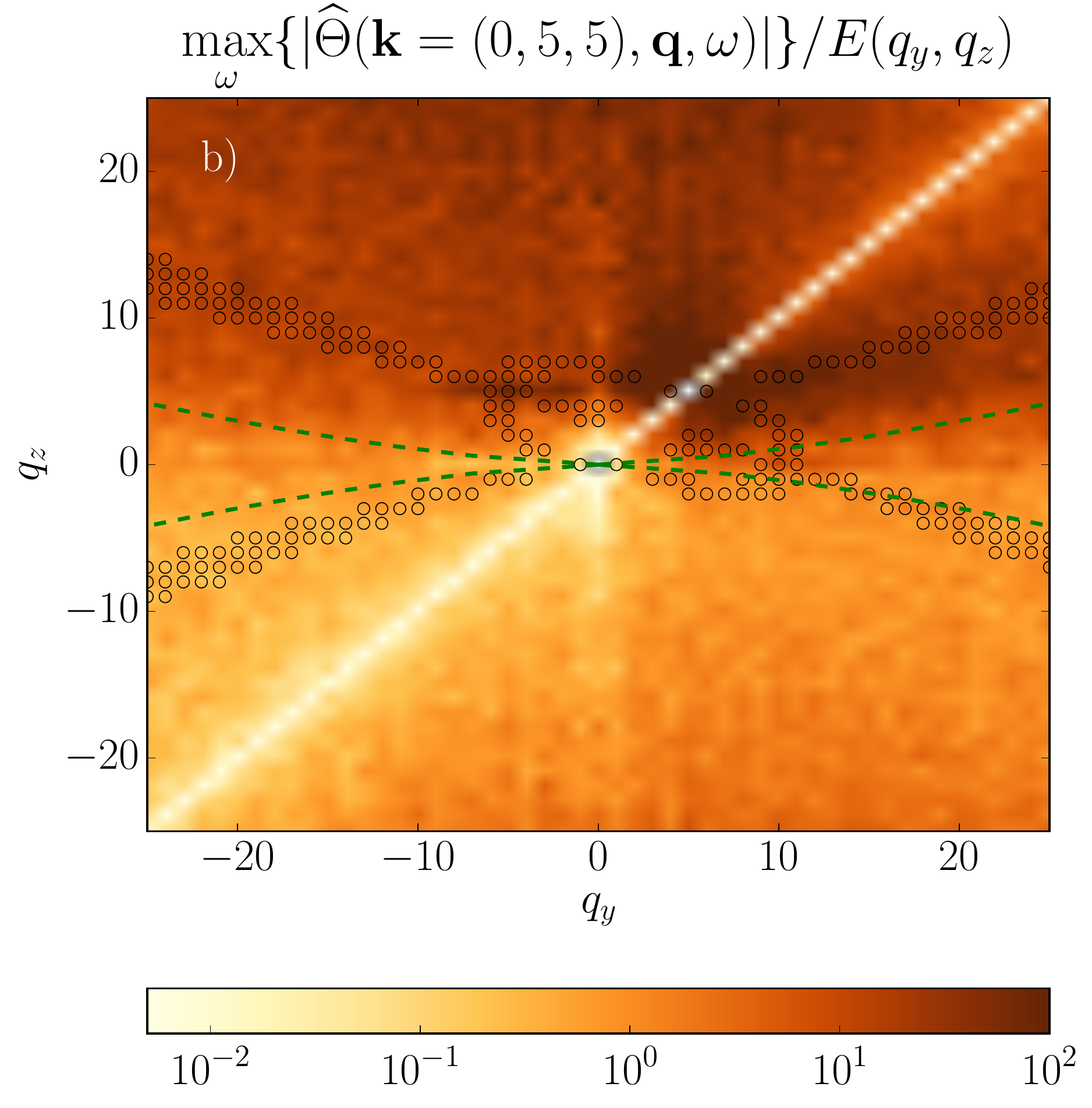}
    \caption{Close up of the geometric distribution of the peak value
      of the normalised contribution function for each triad, given by 
      $\max_\omega \{\vert \widehat{\Theta} ({\bf k},{\bf q},\omega)
      \vert\} /E(q_y,q_z)$. As in Fig.~\ref{peak_frequency}, in each 
      panel ${\bf k}$ is fixed to
      consider two modes dominated by waves: {\it a)} 
      ${\bf k}=(0,0,8)$, and {\it b)} ${\bf k} = (0,5,5)$. The dashed
      curves represent the modes with $\tau_\omega = \tau_{NL}$, and
      the circles represent the near-resonant modes (according to the
      theoretical prediction) with $\gamma_r<0.1$. In good 
      agreement with wave turbulence theories, normalised resonant 
      triads have large amplitudes, but some near-resonant triads are
      also strong. In panel ${\it b)}$, some of these near-resonant triads
      have non-negligible coupling with slow modes (see the circles
      near $q_y \approx 10$ and $q_z \approx 0$) allowing for energy
      transfer towards these modes.}
    \label{zoom}
\end{figure}

\subsection{Contribution functions}

We now present our analysis of the behaviour of the contribution
function $\Theta$. Figure \ref{thetaw} shows the value of 
$\vert \widehat{\Theta}({\bf k}=(0,0,8),{\bf q},\omega)\vert$ as a
function of $\omega$, for four different values of ${\bf q}$ (i.e.,
for four different triads). All of them peak at 
$\omega_0 = \omega({\bf k})$, which is the wave frequency of the mode
${\bf k}$. This was checked for other values of ${\bf k}$ as well, and
a peak in the corresponding wave frequency was observed in all
cases except for the modes with $k_\parallel \approx 0$ (i.e., the
slow modes), for which no discernible peak is present. Moreover, and
in spite of the presence of a peak for modes with $k_\parallel > 0$, 
it is worth noting that the width of the peak depends strongly on the
nature of the other modes in the triad. While interactions with other
wave modes have most of the power in the peak and are well tuned
(i.e., the peak is relatively narrow), interactions with slow modes
can have large amplitudes but with a broad spectrum.

This gives a direct way to identify not only the strength of a given
triad, but also to measure how resonant the triad is, as more resonant
triads are expected to result in a sharper spectrum of 
$ \widehat{\Theta}({\bf k},{\bf q},\omega)$ per virtue of 
Eq.~(\ref{eq:delta}). Therefore, to simplify the analysis, we can 
focus on a few modes ${\bf k}$, explore all available values of 
${\bf q}$ on a triad with ${\bf k}$, and look only at the the maximum
value of $\widehat{\Theta}$ (for all $\omega$) and on the relative
width of the maximum (i.e., on how well tuned the interaction is
around $\omega_0$).

In Fig.~\ref{peak_contribution} we show 
$\max_\omega \{\vert\widehat{\Theta} ({\bf k},{\bf q},\omega) \vert\}$
for two modes ${\bf k}=(0,0,8)$ and $(0,5,5)$, as a function of all
possible values of ${\bf q}$ in the $(0,q_y,q_z)$ plane. The
anisotropic nature of rotating turbulence makes a stellar apparition
here, as the distribution of values is clearly influenced by it. The
result indicates that triads which are elongated along the horizontal
direction have larger amplitudes, which is compatible with the
prediction that energy tends to go towards the slow modes (with 
$q_z\approx 0$) as discussed in \cite{Waleffe92}. Indeed, the triads
with larger amplitudes are located in a horizontal band within $-k_z
\lesssim q_z \lesssim k_z$. There are also strong triads that couple
the ${\bf k}$ mode with modes with larger vertical wavenumber (i.e.,
triads in the horizontal bands $k_z \lesssim |q_z| \lesssim 2 k_z$)
which are compatible with an anisotropic transfer of a fraction of
the energy towards larger wave numbers (i.e., smaller
scales). Finally, collinear modes (i.e., modes with ${\bf q} = \alpha
{\bf k}$) make no contribution to the triads as a result of the
incompressibility of the fluid.

\begin{figure}
    \centering
    \includegraphics[width=6.5cm]{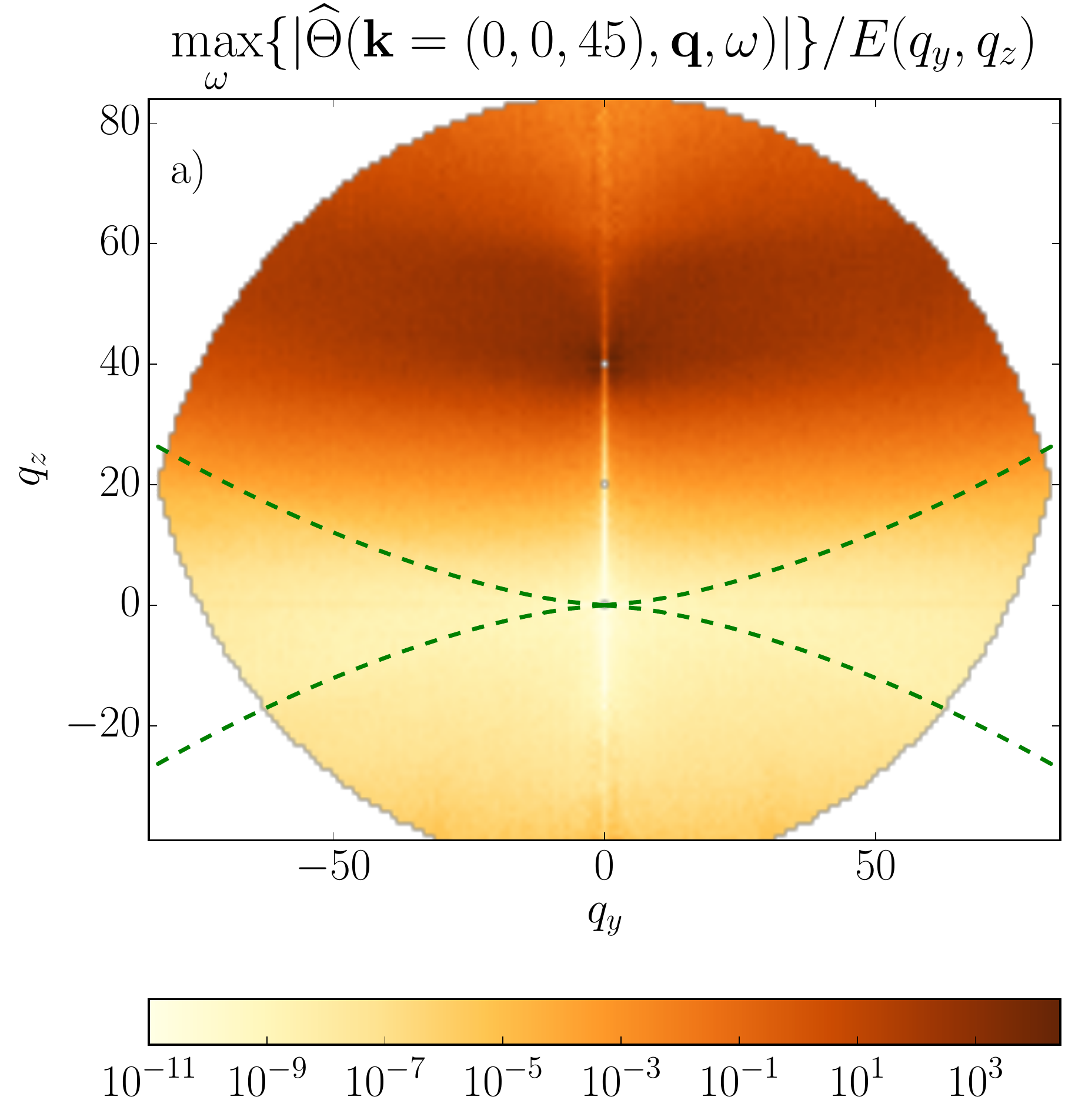}
    \includegraphics[width=6.5cm]{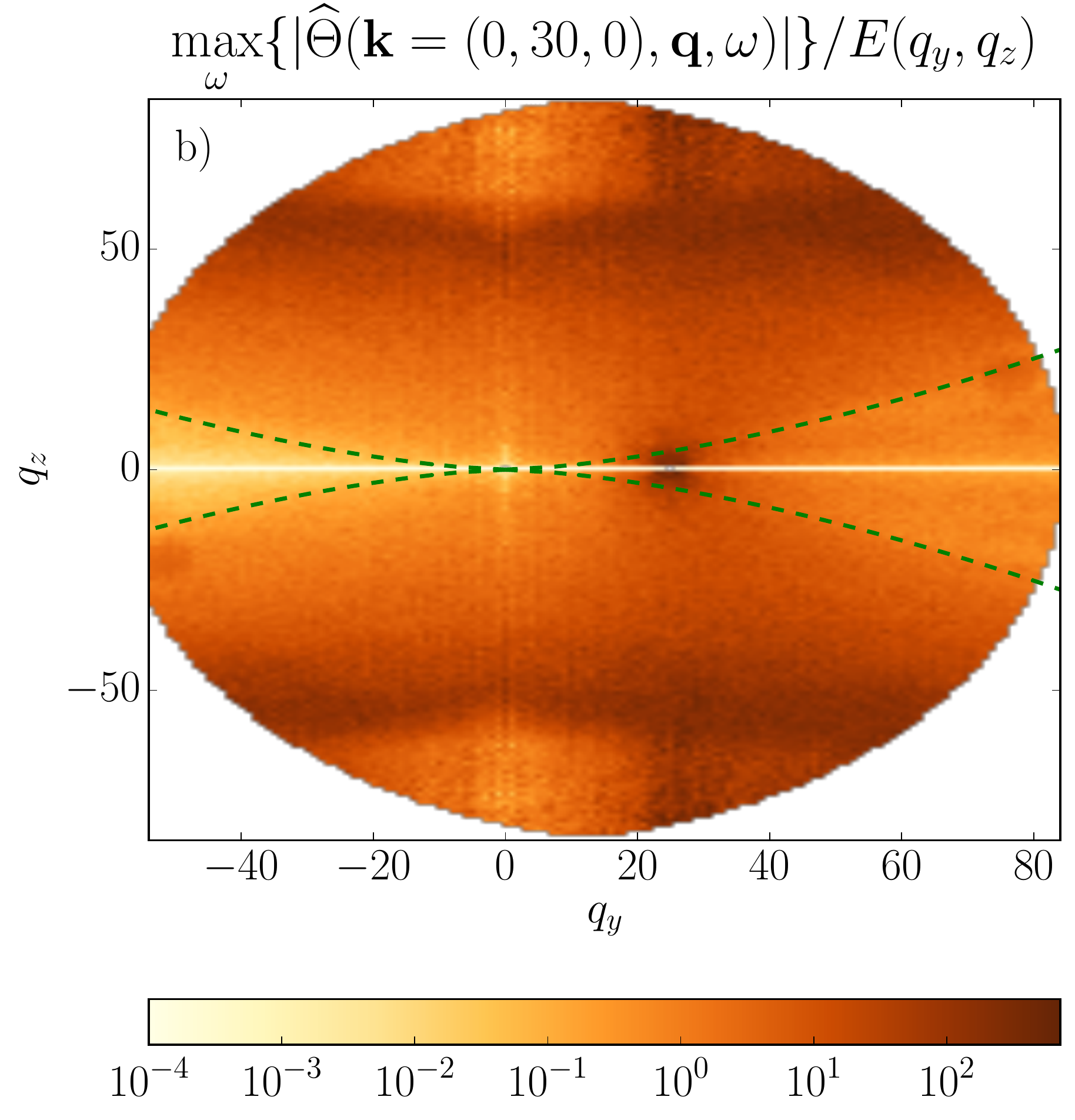}
    \caption{Peak values of the normalised contribution function for
      each triad, given by
      $\max_\omega \{\vert \widehat{\Theta} ({\bf k},{\bf q},\omega)
      \vert\} /E(q_y,q_z)$. The panels correspond to two fixed values 
      of ${\bf k}$: {\it a)} a small-scale ``fast'' mode 
      ${\bf k}=(0,0,45)$, and {\it b)} a small-scale ``slow'' mode 
      ${\bf k} = (0,30,0)$. The dashed curves
      represent the modes with $\tau_\omega = \tau_{NL}$. The role
      played by resonant and near-resonant interactions in these cases
      is less clear.}
    \label{peak_frequency_nowave}
\end{figure}

\begin{figure}
    \centering
    \includegraphics[width=6.5cm]{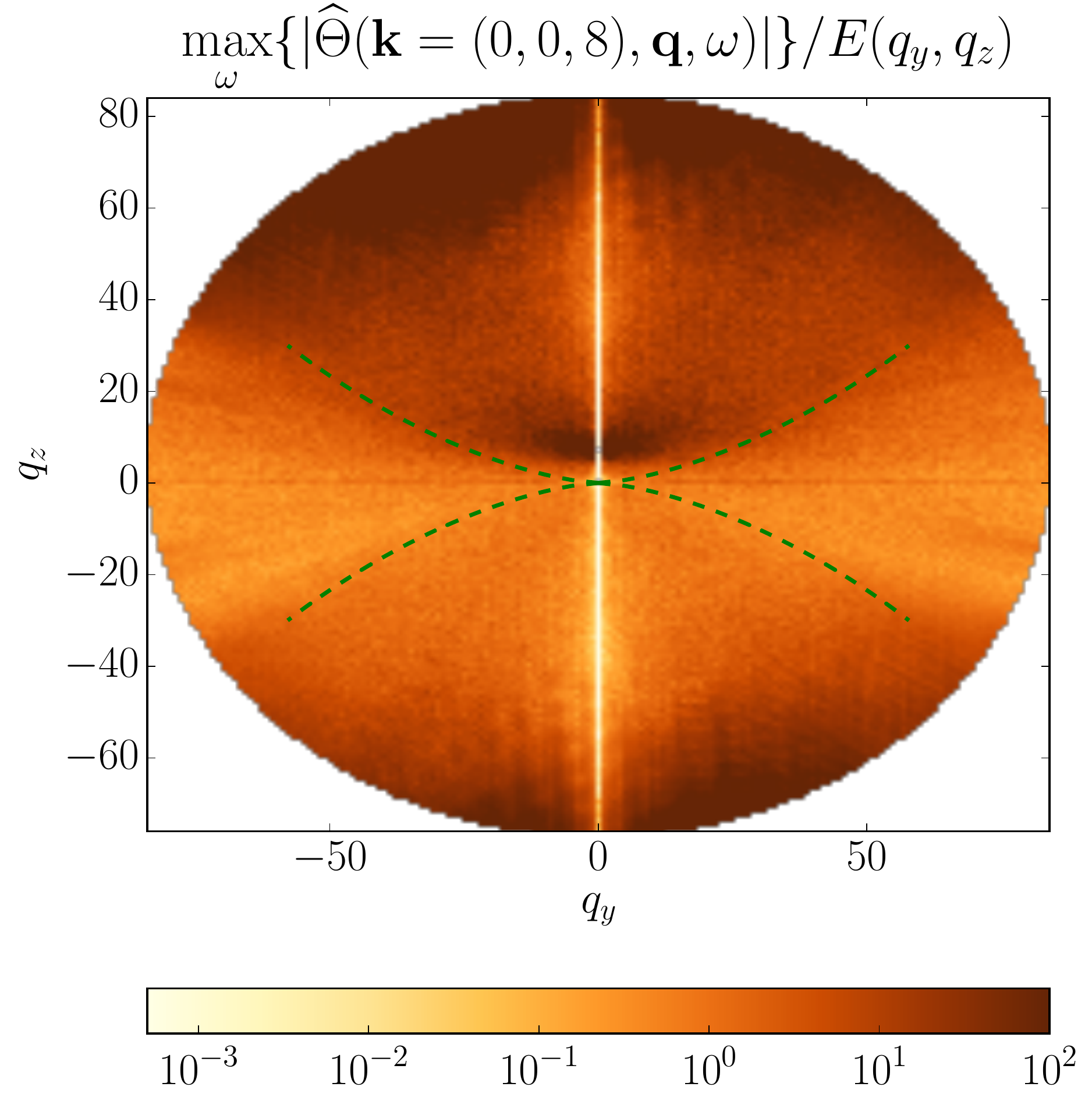}
    \caption{Peak values of the normalised contribution function for
      each triad, 
      $\max_\omega \{\vert \widehat{\Theta} ({\bf k}=(0,0,8),{\bf
        q},\omega) \vert\}/E(q_y,q_z)$, for the simulation 
      with weaker rotation (i.e., larger Rossby
      number). Results are similar to the case with stronger rotation, 
      although the contrast in intensity between triads involving
      fast and slow waves is less clear.}
    \label{peak_frequency_rossby}
\end{figure}

\subsection{Normalised contribution functions}

Having said this, it can be argued that the strongest triads correspond
to modes with $q_z \approx 0$ only as a result of the anisotropic
energy spectrum shown in Fig.~\ref{ekk}: the modes with small vertical
wavenumber have most of the energy, and as a result triads involving
those modes will have larger amplitudes. Therefor, we can normalise the
triads by the energy of the {\bf q} mode in the triad, i.e., we can
compute 
$\max_\omega\{|\hat{\Theta}({\bf k},{\bf q}, \omega)|\}/E({\bf q}).$ If
this is done, from Eqs.~(\ref{eqgamt}) and (\ref{ekwtheta}) the
normalised contribution function has units of inverse time (i.e., of
frequency). This time can thus be interpreted as the time scale of the
energy transfer mechanism, as it is often done in turbulence
theories.

We now turn to the analysis of these normalised contribution
functions. In Fig.~\ref{peak_frequency} we show the peak value of the
normalised functions for ${\bf k}=(0,0,8)$ and $(0,5,5)$. Two dashed
curves indicate the modes for which $\tau_\omega = \tau_{NL}$, i.e.,
modes with the eddy turnover time equal to the period of the
waves. Modes respectively above and below the upper and lower curves
have $\tau_\omega < \tau_{NL}$, and are dominated by the waves. Modes
between the two dashed curves have $\tau_\omega > \tau_{NL}$, and are
slow modes dominated by the eddies.

Anisotropic effects are still evident in Fig.~\ref{peak_frequency}
after the normalisation, but the role played in the triads by the
waves starts to become more clear. Although wave modes have
less energy, after normalisation it becomes evident that triadic
interactions of the wave modes ${\bf k}=(0,0,8)$ and $(0,5,5)$ with
other wave modes are relatively stronger than interactions with
slow modes. If the normalised contribution function is interpreted as
an inverse transfer time, it then implies that the transfer between
triads involving waves is faster than triads involving slow modes, and
thus should be preferred for the interaction. This is true even for
modes ${\bf q}$ with $\tau_s<\tau_\omega<\tau_{NL}$, i.e., for modes
with sweeping time faster than the wave period (but with the waves
still dominating over the eddies).

Large amplitudes (or, equivalently, shorter transfer times) can be
seen in Fig.~\ref{peak_frequency} in the vicinity of 
${\bf q} \approx {\bf k}$. Although at first glance it would appear
that this is due to local (in Fourier space) interactions being the
most prominent, closer inspection reveals that it is also due to the
effect of resonances. In Fig.~\ref{zoom} we show a close up of the
normalised contribution functions in Fig.~\ref{peak_frequency} for
small values of $|{\bf q}|$. Circles mark the modes that satisfy the
theoretical near-resonant condition up to a value of
$\gamma_r=0.1$. It is evident that many of the strongest triads
correspond to resonant or near-resonant triads. This is in very good
agreement with wave turbulence theories of rotating turbulence, which
predict that resonant triads should dominate the coupling between
modes \citep{Newell69}. However, this also
explains how the system transfers energy towards slow modes, which are
inaccessible in weak wave turbulence approximations. The data in
Fig.~\ref{zoom} indicates that not only resonant triads are relevant,
but that near-resonant triads play an equally important role (at least
for the case of a periodic flow). Indeed, large amplitudes can be seen
around the circles in Fig.~\ref{zoom}  in a region that is even
broader than the fan corresponding to the condition
$\gamma_r=0.1$. Close observation of  Fig.~\ref{zoom} {\it a)} and 
{\it b)} (as well as the observation of other modes ${\bf k}$ not
shown here) gives rise to the following picture: For ${\bf k} =
(0,0,8)$, resonant and near-resonant interactions couple this mode
with some modes with $q_z/q < k_z/k$, thus allowing an
energy exchange between these modes. Energy can thus be transferred
towards modes with smaller vertical wavenumber, in agreement with the
arguments in \cite{Cambon89} and \cite{Waleffe93}. For 
${\bf k} = (0,5,5)$, the process is repeated, but now some
near-resonant interactions allow for a coupling (and thus a transfer)
with slow modes. This is compatible with observations in
\cite{Smith05}, \cite{Alexakis15}, and \cite{Gallet15}. Moreover, in
Fig.~\ref{zoom} {\it b)} a non-negligible coupling with slow modes can
be observed even for non-resonant triads (see the region between the
two-dashed curves with $q_y>0$), indicating that as energy approaches
the slow modes the role of non-resonant interactions may also become
more relevant.

\begin{figure}
    \centering
    \includegraphics[width=6.5cm]{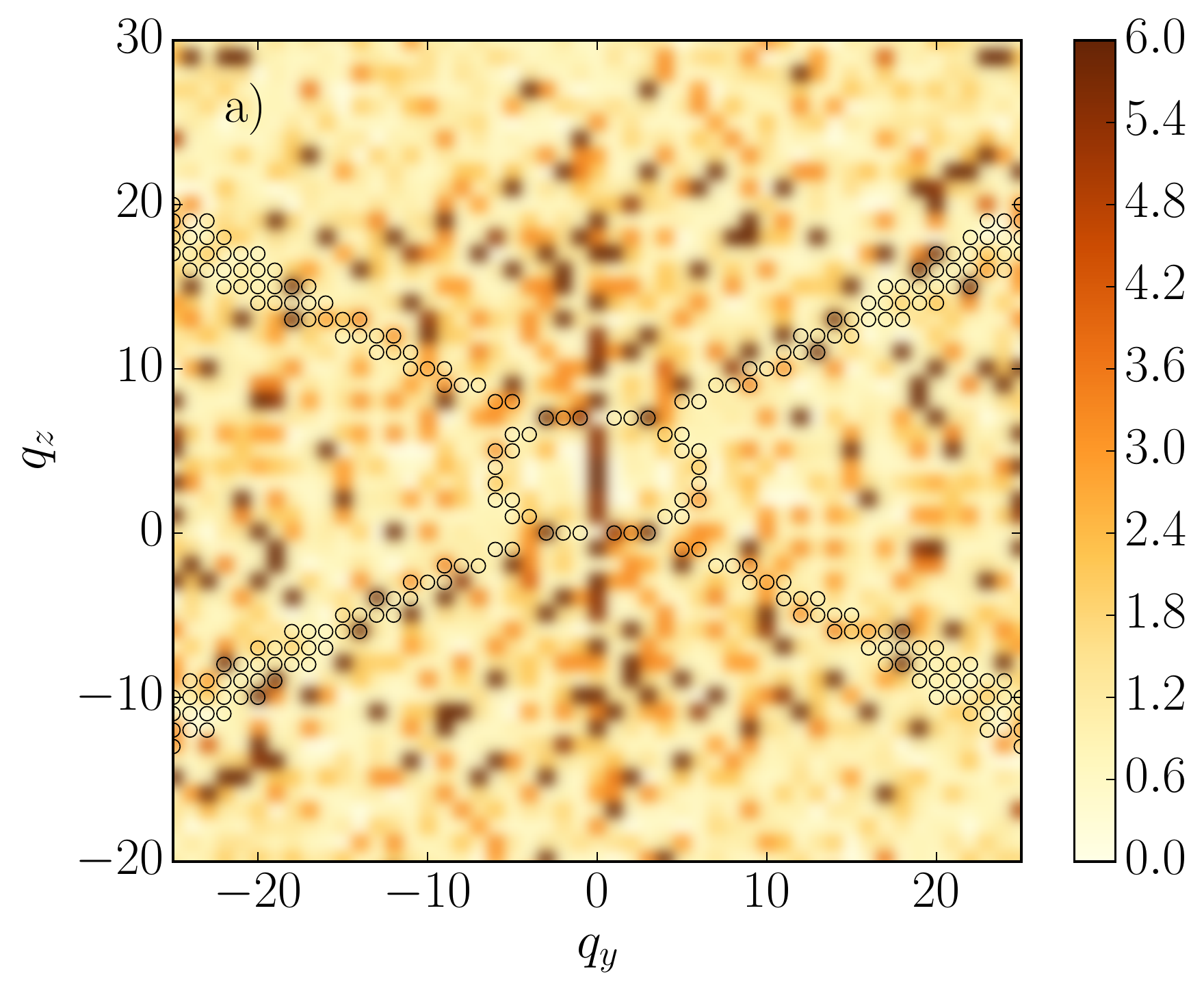}
    \includegraphics[width=6.5cm]{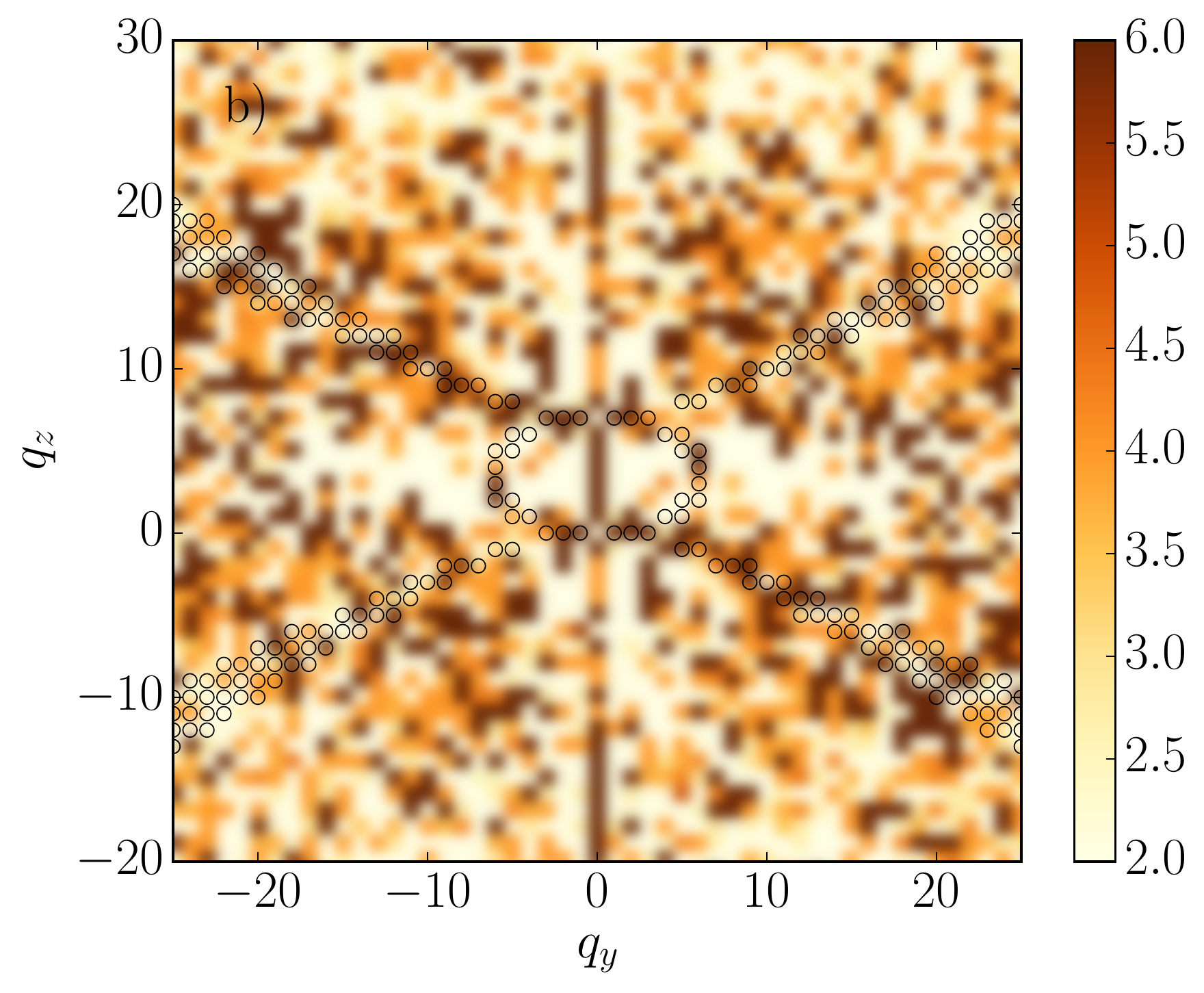}
    \caption{Inverse relative bandwidth (quality factor) of the peak in
      each contribution function, as a function of ${\bf q}$, and for
      fixed ${\bf k} =(0,0,8)$, {\it a)} for the simulation with $Ro
      \approx 0.03$, and {\it b)} for the simulation with $Ro \approx
      0.015$. Larger quality factors correspond to sharper bandwidths
      of the triads relative to their central frequency, and thus to more
      resonant interactions (i.e., the factor quantifies how well tuned a
      triad is). Circles indicate modes which satisfy the theoretical
      near-resonance condition with $\gamma_r<0.1$. A good agreement
      is observed between the theoretical condition and the quality
      factor of the contribution function, specially for the flow with
      smaller Rossby number. Note however that relatively large
      quality factors are observed for branches which are wider than
      the condition $\gamma_r<0.1$, indicating again the important
      role played by near-resonant interactions.}
    \label{tuning}
\end{figure}

\subsection{Comparison with small-scale and slow modes}

To gain more certainty on the effect of the waves in the triadic
interactions, we compare now the previous results with the normalised
contribution function for two modes: a small-scale mode with 
${\bf k}=(0,0,45)$, which is dominated by sweeping (but with the wave
period faster than the eddy turnover time), and a small-scale slow
mode with ${\bf k}=(0,30,0)$ which has zero wave frequency. Here, by
small-scale, we refer to modes with wave numbers significantly larger
than the forced wave number. The resulting normalised 
contribution functions are shown in 
Fig.~\ref{peak_frequency_nowave}. In both cases, the division
given by the curve with $\tau_\omega = \tau_{NL}$ is less evident, and a
superposition of the modes expected 
to be resonant or near-resonant (not shown)
indicates no clear correlation between the strength of the triad and
the theoretical resonant or near-resonant condition. For the mode 
${\bf k}=(0,0,45)$, the normalised contribution function indicates
that coupling is stronger for modes with $q_z \approx k_z$, an effect
associated with the anisotropy of the flow, while the coupling with slow
modes is negligible. The slow mode ${\bf k}=(0,0,45)$ shows more
interesting features. The mode seems to be more strongly coupled with
other slow modes ${\bf q}$ in the vicinity of  ${\bf k}$, and with
modes with large $q_z$ (of the order of $|{\bf k}|$, compatible with
local triadic interactions, although these interactions are
non-resonant).

\subsection{Effect of the Rossby number}

We can also compare the results in the two simulations with different
Rossby number, to quantify the effect of changing the rotation
frequency in the intensity of resonant and near-resonant
triads. As an illustration, Fig.~\ref{peak_frequency_rossby} shows the
geometric distribution of the peak values of the normalised
contribution function for the mode ${\bf k}=(0,0,8)$ in the simulation
with larger Rossby number. The same result as in
Fig.~\ref{peak_frequency} is obtained, but the contrast between modes
above and below the $\tau_\omega=\tau_{NL}$ curve is less
marked. Also, the region of modes dominated by eddies (i.e., of
slow modes) increases as expected, and the boundary between the
two regions indicated by the change in intensity of the triads moves
accordingly. This confirms that the changes in intensity in
Fig.~\ref{peak_frequency} indeed separate triads involving slow and
fast modes, and is also consistent with the behaviour expected in a
rotating flow as the Rossby number is varied.

\subsection{Direct measurement of the resonance level of each triad}

One of the most important implications of the contribution function is
that it allows a direct measurement of how well tuned a triad is,
i.e., of how resonant the interaction between three modes is. This was
already discussed in the context of Fig.~\ref{thetaw}, where we showed
that some triads display a narrower peak around the wave frequency
than others. But now we can put this observation on firmer grounds.

As it follows from Eq.~(\ref{eq:delta}), for a perfectly resonant
triad the contribution function should be a delta distribution
centred around $\omega_0=\omega({\bf k})$. Near-resonant and
non-resonant interactions broaden the peak. This broadening can be
measured using the quality factor 
\begin{equation}
Q = \frac{\omega_0}{\Delta \omega} .
\end{equation}
In other words, the $Q$ factor is the inverse relative width of the
peak of $\widehat{\Theta}({\bf k},{\bf q},\omega)$ as a function
of the frequency. We estimate $\Delta \omega$ by calculating the
  width of the peak in the spectrum (see Fig.~\ref{thetaw}) at half
  the amplitude of the maximum value. In the theory of resonators, the
$Q$ factor is often interpreted as the ratio of the energy stored to
the energy lost in a system. In our case, the larger the $Q$ factor
the more resonant the triad is, and the less energy of the mode 
${\bf k}$ is lost (i.e., given) to non-resonant modes.

Figure \ref{tuning} shows this quality factor for each contribution
function for fixed ${\bf k}=(0,0,8)$ and for all possible values of
${\bf q} = (0,q_y,q_z)$, for the two simulations with different Rossby
numbers. Superimposed to the quality factor, circles mark the modes that
satisfy the theoretical near-resonant condition up to a value of
$\gamma_r=0.1$. In particular for the simulation with smaller Rossby
number, triads with large quality factors (i.e., well tuned triads)
more or less coincide with triads with small $\gamma_r$, specially for
the branches in the upper-left and lower-right quadrants on the figure
on the right in Fig.~\ref{tuning}. In other words, the quality factor
defined above gives a good measure of how resonant a triad
is. Moreover, three features in Fig.~\ref{tuning} are worth
emphasising: First, the $Q$ factor has maximum values $\approx 6$ 
which is of the order of, although a bit smaller than, what is
often found in electrical or mechanical resonators. In other words,
even resonant triads display broadband peaks in the frequency
spectrum. Second, the area covered by triads with the largest values
of $Q$ is relatively larger than the area corresponding  to the
circles with $\gamma_r<0.1$, confirming the importance of interactions
that are even marginally near-resonant. The scaling of this
  behaviour with Rossby number and domain size can be important for
  theories of rotating turbulence in infinite domains, in which the
  behavior of the resonant and slow manifolds, as well as the
  relevance of near-resonances, are unclear
  \citep{Cambon04,Chen05,Bourouiba08}. Third, there are some modes
with relatively large $Q$ values (compared with the mean amplitude of
$Q$ for all modes) that do not correspond to resonant or near-resonant
modes in wave turbulence theory. Most notably, some of these modes are
modes with small $|{\bf q}|$ and lying in the region of the slow modes.

\section{Conclusions}
\label{conclusions}

One of the central problems in turbulence theory involves the
understanding of how modes interact non-linearly, specially in systems
with restitutive forces for which eddies and waves can coexist, and
for which resonances can strongly affect the nonlinear triadic
interactions. In these problems, a direct investigation of how each
triad of modes contributes to the overall dynamics is quite cumbersome
and complicated. To tackle this problem we have derived a
contribution function that characterises the spatio-temporal behaviour
of each triad, and more importantly, their contribution to the energy
transfer and to the spatio-temporal spectrum of the turbulent
flow.

We used this function to study the case of rotating turbulence, in
which eddies coexist with inertial waves, and where triadic resonant
interactions are expected to be dominant \citep{Newell69},
transferring their energy preferentially towards modes with small vertical
wave numbers \citep{Cambon89,Waleffe93}. However, this picture fails to
explain how energy continues to be transferred anisotropically to
``slow'' two-dimensional modes, as the wave turbulence approximation
breaks down in the vicinity of those modes. Previous results in
simulations at low resolution or in truncated systems
\citep{Chen05,Smith05} indicate that near-resonant triads can be
responsible for this latter transfer, but it is still unclear whether
these interactions remain to be relevant as the turbulence level is 
increased, or as the Rossby number is decreased. Some recent results
suggest this to be the case \citep{Alexakis15,Gallet15}.

We computed the contribution function for a large number of triads in
two simulations of rotating turbulence at spatial resolutions of
$512^3$ grid points. As the contribution function is a third-order
time-correlation function for each Fourier triad, this requires a
massive analysis of spatio-temporal data. The main results show
or confirm that:
(1) For ``wave'' modes with $\tau_\omega<\tau_s<\tau_{NL}$ (i.e.,
modes for which the wave period is faster than the sweeping time and
the eddy turnover time), the coupling between triads is strongly 
anisotropic. Triads which are elongated along the horizontal direction
have larger amplitudes, a result which is compatible with the
prediction that energy tends to go towards modes with smaller parallel
wave numbers. This result is also in agreement with the 
proposed mechanism of parametric instability, which was obtained for
isolated triads \citep{Waleffe93}, while the data analysis presented
here considers the system with all possible couplings between the
triads.
(2) After normalising the triads by the energy in one of modes, it is
found that the transfer between triads involving wave modes is faster
than the transfer between triads that couple the wave mode with a slow
mode, and thus the former can be expected to be the preferred ones for
interactions, also in agreement with predictions from wave turbulence
theory.
(3) However, near-resonant and non-resonant interactions are
non-negligible, and couple the wave modes to slow modes, thus allowing
for energy transfer into that region of spectral space.
(4) The contribution function is peaked around the frequency of each
mode, and thus can be used to define a quality factor $Q$ that
measures how resonant a triad is. While resonant triads are compatible
with relatively larger values of the $Q$ factor, the analysis shows that
some marginally near-resonant and non-resonant triads also display
tuning with the wave frequency and are such that can couple fast and
slow modes. Further studies of this result can be importat for
  theories of rotating turbulence in infinite domains, in which the
  nature and coupling of the modes in the slow and in the resonant
  manifolds with the rest of the modes is a matter of debate.
(6) For modes for which $\tau_\omega$ is larger than $\tau_s$ or
$\tau_{NL}$, the relevance of resonant and near-resonant triads
decreases rapidly.
(7) Finally, varying the Rossby number qualitatively preserves these
results, at least in the short range of values considered here.

These results are in agreement with major theoretical predictions for
the behaviour of nonlinear interactions in rotating turbulence, as
mentioned above, and can shed some light on the recent results
concerning the behaviour or two-dimensional modes for very small
Rossby numbers. In this context, an obvious shortcoming of the present
study is the lack of a parametric study of the behaviour of the triads
for even smaller Rossby numbers, or for larger Reynolds numbers. The
need to properly resolve in time the fastest waves and to store with
high time cadence the data, to then perform the spatio-temporal
analysis of each triad, precludes for the moment studies with faster
rotation or with larger spatial resolution. However, we believe that
the results presented here can be useful to quantitatively assess the
relevance of resonant, near-resonant, and non-resonant triads at
moderate Rossby numbers. Also, the formalism presented here can be
extended to analyse other systems in which resonant interactions 
are also believed to play a central role (see, e.g., the recent
studies by \citet{Aubourg15,Aubourg16}).

\begin{acknowledgements}
    The authors acknowledge support from Grant Nos. PIP
    11220090100825, UBACYT 20020110200359, and PICT 2011-1529.
    PCdL thanks Luca Biferale for fruitful discussions and comments.
\end{acknowledgements}

\bibliographystyle{jfm}
\bibliography{ms}

\begin{thebibliography}{54}
\expandafter\ifx\csname natexlab\endcsname\relax\def\natexlab#1{#1}\fi

\bibitem[Alexakis(2015)]{Alexakis15}
{\sc Alexakis, A.} 2015 Rotating {Taylor}-{Green} flow. {\em J.\ Fluid Mech.\/}
  {\bf 769}, 46--78.

\bibitem[Aluie \& Eyink(2009)]{Aluie09}
{\sc Aluie, H. \& Eyink, G.~L.} 2009 Localness of energy cascade in
  hydrodynamic turbulence. {I}{I}. sharp spectral filter. {\em Phys.\ Fluids\/}
  {\bf 21}, 115108.

\bibitem[Aubourg \& Mordant(2015)]{Aubourg15}
{\sc Aubourg, Q. \& Mordant, N.} 2015 Nonlocal {Resonances} in {Weak}
  {Turbulence} of {Gravity}-{Capillary} {Waves}. {\em Phys.\ Rev.\ Lett.\/}
  {\bf 114}, 144501.

\bibitem[Aubourg \& Mordant(2016)]{Aubourg16}
{\sc Aubourg, Q. \& Mordant, N.} 2016 Investigation of resonances in
  gravity-capillary wave turbulence. {\em Phys.\ Rev.\ Fluids\/} {\bf 1},
  023701.

\bibitem[Bellet {\em et~al.\/}(2006)Bellet, Godeferd, Scott \&
  Cambon]{Bellet06}
{\sc Bellet, F., Godeferd, F.~S., Scott, J.~F. \& Cambon, C.} 2006 Wave
  turbulence in rapidly rotating flows. {\em J.\ Fluid Mech.\/} {\bf 562},
  83--121.

\bibitem[Bewley {\em et~al.\/}(2007)Bewley, Lathrop, Maas \&
  Sreenivasan]{Bewley07}
{\sc Bewley, G.~P., Lathrop, D.~P., Maas, L. R.~M. \& Sreenivasan, K.~R.} 2007
  Inertial waves in rotating grid turbulence. {\em Phys.\ Fluids\/} {\bf
  19}~(7), 071701.

\bibitem[Biferale {\em et~al.\/}(2013)Biferale, Musacchio \&
  Toschi]{Biferale13}
{\sc Biferale, L., Musacchio, S. \& Toschi, F.} 2013 Split energy-helicity
  cascades in three-dimensional homogeneous and isotropic turbulence. {\em J.\
  Fluid Mech.\/} {\bf 730}, 309--327.

\bibitem[Bokhoven {\em et~al.\/}(2008)Bokhoven, Cambon, Liechtenstein, Godeferd
  \& Clercx]{Bokhoven08}
{\sc Bokhoven, L. J. A.~van, Cambon, C., Liechtenstein, L., Godeferd, F.~S. \&
  Clercx, H. J.~H.} 2008 Refined vorticity statistics of decaying rotating
  three-dimensional turbulence. {\em J.\ Turbul.\/} {\bf 9}, N6.

\bibitem[Bordes {\em et~al.\/}(2012)Bordes, Moisy, Dauxois \& Cortet]{Bordes12}
{\sc Bordes, G., Moisy, F., Dauxois, T. \& Cortet, P.-P.} 2012 Experimental
  evidence of a triadic resonance of plane inertial waves in a rotating fluid.
  {\em Phys.\ Fluids\/} {\bf 24}~(1), 014105.

\bibitem[Bourouiba(2008)]{Bourouiba08}
{\sc Bourouiba, Lydia} 2008 Discreteness and resolution effects in rapidly
  rotating turbulence. {\em Phys.\ Rev.\ E\/} {\bf 78}~(5), 056309.

\bibitem[Cambon \& Jacquin(1989)]{Cambon89}
{\sc Cambon, C. \& Jacquin, L.} 1989 Spectral approach to non-isotropic
  turbulence subjected to rotation. {\em J.\ Fluid Mech.\/} {\bf 202},
  295--317.

\bibitem[Cambon {\em et~al.\/}(1997)Cambon, Mansour \& Godeferd]{Cambon97}
{\sc Cambon, C., Mansour, N.~N. \& Godeferd, F.~S.} 1997 Energy transfer in
  rotating turbulence. {\em J.\ Fluid Mech.\/} {\bf 337}, 303--332.

\bibitem[Cambon {\em et~al.\/}(2004)Cambon, Rubinstein \& Godeferd]{Cambon04}
{\sc Cambon, C., Rubinstein, R. \& Godeferd, F.~S.} 2004 Advances in wave
  turbulence: rapidly rotating flows. {\em New J.\ Phys.\/} {\bf 6}, 73.

\bibitem[Campagne {\em et~al.\/}(2015)Campagne, Gallet, Moisy \&
  Cortet]{Campagne15}
{\sc Campagne, A., Gallet, B., Moisy, F. \& Cortet, P.-P.} 2015 Disentangling
  inertial waves from eddy turbulence in a forced rotating--turbulence
  experiment. {\em Phys.\ Rev.\ E\/} {\bf 91}~(4), 043016.

\bibitem[Chen {\em et~al.\/}(2005)Chen, Chen, Eyink \& Holm]{Chen05}
{\sc Chen, Q., Chen, S., Eyink, G.~L. \& Holm, D.~D.} 2005 Resonant
  interactions in rotating homogeneous three-dimensional turbulence. {\em J.\
  Fluid Mech.\/} {\bf 542}, 139--164.

\bibitem[Chen \& Kraichnan(1989)]{Chen89}
{\sc Chen, S. \& Kraichnan, R.~H.} 1989 Sweeping decorrelation in isotropic
  turbulence. {\em Phys.\ Fluids A\/} {\bf 1}~(12), 2019.

\bibitem[Cheung \& Zaki(2014)]{Cheung14}
{\sc Cheung, L.~C. \& Zaki, T.~A.} 2014 An exact representation of the
  nonlinear triad interaction terms in spectral space. {\em J.\ Fluid Mech.\/}
  {\bf 748}, 175--188.

\bibitem[Domaradzki \& Rogallo(1990)]{Domaradzki90a}
{\sc Domaradzki, J.~Andrzej \& Rogallo, Robert~S.} 1990 Local energy transfer
  and nonlocal interactions in homogeneous, isotropic turbulence. {\em Phys.\
  Fluids A\/} {\bf 2}~(3), 413--426.

\bibitem[Eyink \& Aluie(2009)]{Eyink09}
{\sc Eyink, G.~L. \& Aluie, H.} 2009 Localness of energy cascade in
  hydrodynamic turbulence. {I}. smooth coarse graining. {\em Phys.\ Fluids\/}
  {\bf 21}, 115107.

\bibitem[Favier {\em et~al.\/}(2010)Favier, Godeferd \& Cambon]{Favier10}
{\sc Favier, B., Godeferd, F.~S. \& Cambon, C.} 2010 On space and time
  correlations of isotropic and rotating turbulence. {\em Phys.\ Fluids\/} {\bf
  22}~(1), 015101.

\bibitem[Gallet(2015)]{Gallet15}
{\sc Gallet, B.} 2015 Exact two-dimensionalization of rapidly rotating
  large-{Reynolds}-number flows. {\em J.\ Fluid Mech.\/} {\bf 783}, 412--447.

\bibitem[Galtier(2003)]{Galtier03}
{\sc Galtier, S.} 2003 Weak inertial-wave turbulence theory. {\em Phys.\ Rev.\
  E\/} {\bf 68}, 015301.

\bibitem[G\'omez {\em et~al.\/}(2005)G\'omez, Mininni \& P.]{Gomez05}
{\sc G\'omez, D.~O., Mininni, P.~D. \& P., Dmitruk.} 2005 M{H}{D} simulations
  and astrophysical applications. {\em Adv.\ Sp.\ Res.\/} {\bf 35}, 899--907.

\bibitem[Haudin {\em et~al.\/}(2016)Haudin, Cazaubiel, Deike, Jamin, Falcon \&
  Berhanu]{Haudin16}
{\sc Haudin, F., Cazaubiel, A., Deike, L., Jamin, T., Falcon, E. \& Berhanu,
  M.} 2016 Experimental study of three-wave interactions among
  capillary-gravity surface waves. {\em Phys.\ Rev.\ E\/} {\bf 93}~(4), 043110.

\bibitem[Hernandez-Duenas {\em et~al.\/}(2014)Hernandez-Duenas, Smith \&
  Stechmann]{Hernandez-duenas14}
{\sc Hernandez-Duenas, Gerardo, Smith, Leslie~M. \& Stechmann, Samuel~N.} 2014
  Investigation of {Boussinesq} dynamics using intermediate models based on
  wave–vortical interactions. {\em J.\ Fluid Mech.\/} {\bf 747}, 247--287.

\bibitem[Horne \& Mininni(2013)]{Horne13}
{\sc Horne, E. \& Mininni, P.~D.} 2013 Sign cancellation and scaling in the
  vertical component of velocity and vorticity in rotating turbulence. {\em
  Phys.\ Rev.\ E\/} {\bf 88}~(1), 013011.

\bibitem[Kraichnan(1958)]{Kraichnan58}
{\sc Kraichnan, R.~H.} 1958 Irreversible statistical mechanics of
  incompressible hydromagnetic turbulence. {\em Phys.\ Rev.\/} {\bf 109}~(5),
  1407--1422.

\bibitem[Lamriben {\em et~al.\/}(2011)Lamriben, Cortet, Moisy \&
  Maas]{Lamriben11}
{\sc Lamriben, C., Cortet, P.-P., Moisy, F. \& Maas, L. R.~M.} 2011 Excitation
  of inertial modes in a closed grid turbulence experiment under rotation. {\em
  Phys.\ Fluids\/} {\bf 23}~(1), 015102.

\bibitem[Lee(1975)]{Lee75}
{\sc Lee, J.} 1975 The triad‐interaction representation of homogeneous
  turbulence. {\em J.\ Math.\ Phys.\/} {\bf 16}~(7), 1359--1366.

\bibitem[Clark~di Leoni {\em et~al.\/}(2015)Clark~di Leoni, Cobelli \&
  Mininni]{Clark15b}
{\sc Clark~di Leoni, P., Cobelli, P.~J. \& Mininni, P.~D.} 2015 The
  spatio-temporal spectrum of turbulent flows. {\em Euro.\ Phys.\ J.\ E\/} {\bf
  38}~(12), 1--10.

\bibitem[Clark~di Leoni {\em et~al.\/}(2014)Clark~di Leoni, Cobelli, Mininni,
  Dmitruk \& Matthaeus]{Clark14}
{\sc Clark~di Leoni, P., Cobelli, P.~J., Mininni, P.~D., Dmitruk, P. \&
  Matthaeus, W.~H.} 2014 Quantification of the strength of inertial waves in a
  rotating turbulent flow. {\em Phys.\ Fluids\/} {\bf 26}~(3), 035106.

\bibitem[Mininni(2011)]{Mininni11b}
{\sc Mininni, P.~D.} 2011 Scale interactions in magnetohydrodynamic turbulence.
  {\em Annu.\ Rev.\ Fluid Mech.\/} {\bf 43}~(1), 377--397.

\bibitem[Mininni {\em et~al.\/}(2006)Mininni, Alexakis \& Pouquet]{Mininni06}
{\sc Mininni, P.~D., Alexakis, A. \& Pouquet, A.} 2006 Large-scale flow
  effects, energy transfer, and self-similarity on turbulence. {\em Phys.\
  Rev.\ E\/} {\bf 74}~(1).

\bibitem[Mininni {\em et~al.\/}(2008)Mininni, Alexakis \& Pouquet]{Mininni08}
{\sc Mininni, P.~D., Alexakis, A. \& Pouquet, A.} 2008 Nonlocal interactions in
  hydrodynamic turbulence at high {R}eynolds numbers: The slow emergence of
  scaling laws. {\em Phys.\ Rev.\ E\/} {\bf 77}~(3).

\bibitem[Mininni {\em et~al.\/}(2012)Mininni, Rosenberg \& Pouquet]{Mininni12}
{\sc Mininni, P.~D., Rosenberg, D. \& Pouquet, A.} 2012 Isotropization at small
  scales of rotating helically driven turbulence. {\em J.\ Fluid Mech.\/} {\bf
  699}, 263--279.

\bibitem[Mininni {\em et~al.\/}(2011)Mininni, Rosenberg, Reddy \&
  Pouquet]{Mininni11}
{\sc Mininni, P.~D., Rosenberg, D., Reddy, R. \& Pouquet, A.} 2011 A hybrid
  {M}{P}{I}--{O}pen{M}{P} scheme for scalable parallel pseudospectral
  computations for fluid turbulence. {\em Parallel Computing\/} {\bf 37},
  316--326.

\bibitem[Moffatt(2014)]{Moffatt14}
{\sc Moffatt, H.~K.} 2014 Note on the triad interactions of homogeneous
  turbulence. {\em J.\ Fluid Mech.\/} {\bf 741}.

\bibitem[M\"uller \& Thiele(2007)]{Muller07}
{\sc M\"uller, W.-C. \& Thiele, M.} 2007 Scaling and energy transfer in
  rotating turbulence. {\em Europhys.\ Lett.\/} {\bf 77}, 34003.

\bibitem[Nazarenko(2011)]{Nazarenko}
{\sc Nazarenko, S.} 2011 {\em Wave Turbulence\/}, 2011th edn. Springer.

\bibitem[Nazarenko \& Schekochihin(2011)]{Nazarenko11}
{\sc Nazarenko, S.~V. \& Schekochihin, A.~A.} 2011 Critical balance in
  magnetohydrodynamic, rotating and stratified turbulence: towards a universal
  scaling conjecture. {\em J.\ Fluid Mech.\/} {\bf 677}, 134--153.

\bibitem[Newell(1969)]{Newell69}
{\sc Newell, A.~C.} 1969 Rossby wave packet interactions. {\em J.\ Fluid
  Mech.\/} {\bf 35}~(02), 255--271.

\bibitem[Newell \& Rumpf(2011)]{Newell11}
{\sc Newell, A.~C. \& Rumpf, B.} 2011 Wave turbulence. {\em Annu.\ Rev.\ Fluid
  Mech.\/} {\bf 43}~(1), 59--78.

\bibitem[Pouquet \& Mininni(2010)]{Pouquet10}
{\sc Pouquet, A. \& Mininni, P.~D.} 2010 The interplay between helicity and
  rotation in turbulence: implications for scaling laws and small-scale
  dynamics. {\em Phil.\ Trans.\ Royal Soc.\ London A\/} {\bf 368}, 1635--1662.

\bibitem[Remmel {\em et~al.\/}(2010)Remmel, Sukhatme \& Smith]{Remmel10}
{\sc Remmel, Mark, Sukhatme, Jai \& Smith, Leslie~M.} 2010 Nonlinear
  inertia-gravity wave-mode interactions in three dimensional rotating
  stratified flows. {\em Commun. Math. Sci.\/} {\bf 8}~(2), 357--376.

\bibitem[Rieutord {\em et~al.\/}(2012)Rieutord, Triana, Zimmerman \&
  Lathrop]{Rieutord12}
{\sc Rieutord, M., Triana, S.~A., Zimmerman, D.~S. \& Lathrop, D.~P.} 2012
  Excitation of inertial modes in an experimental spherical {Couette} flow.
  {\em Phys.\ Rev.\ E\/} {\bf 86}~(2), 026304.

\bibitem[Sen {\em et~al.\/}(2012)Sen, Mininni, Rosenberg \& Pouquet]{Sen12}
{\sc Sen, A., Mininni, P.~D., Rosenberg, D. \& Pouquet, A.} 2012 Anisotropy and
  nonuniversality in scaling laws of the large-scale energy spectrum in
  rotating turbulence. {\em Phys.\ Rev.\ E\/} {\bf 86}, 036319.

\bibitem[Servidio {\em et~al.\/}(2011)Servidio, Carbone, Dmitruk \&
  Matthaeus]{Servidio11}
{\sc Servidio, S., Carbone, V., Dmitruk, P. \& Matthaeus, W.~H.} 2011 Time
  decorrelation in isotropic magnetohydrodynamic turbulence. {\em Europhys.\
  Lett.\/} {\bf 96}~(5), 55003.

\bibitem[Smith \& Lee(2005)]{Smith05}
{\sc Smith, L.~M. \& Lee, Y.} 2005 On near resonances and symmetry breaking in
  forced rotating flows at moderate {Rossby} number. {\em J.\ Fluid Mech.\/}
  {\bf 535}, 111--142.

\bibitem[Staplehurst {\em et~al.\/}(2008)Staplehurst, Davidson \&
  Dalziel]{Staplehurst08}
{\sc Staplehurst, P.~J., Davidson, P.~A. \& Dalziel, S.~B.} 2008 Structure
  formation in homogeneous freely decaying rotating turbulence. {\em J.\ Fluid
  Mech.\/} {\bf 598}, 81--105.

\bibitem[Waleffe(1992)]{Waleffe92}
{\sc Waleffe, F.} 1992 The nature of triad interactions in homogeneous
  turbulence. {\em Phys.\ Fluids A\/} {\bf 4}~(2), 350--363.

\bibitem[Waleffe(1993)]{Waleffe93}
{\sc Waleffe, F.} 1993 Inertial transfers in the helical decomposition. {\em
  Phys.\ Fluids A\/} {\bf 5}~(3), 677.

\bibitem[Yarom \& Sharon(2014)]{Yarom14}
{\sc Yarom, E. \& Sharon, E.} 2014 Experimental observation of steady inertial
  wave turbulence in deep rotating flows. {\em Nature Phys.\/} {\bf 10}~(7),
  510--514.

\bibitem[Zakharov {\em et~al.\/}(1992)Zakharov, Lvov \& Falkovic]{Zakharov}
{\sc Zakharov, V.~E., Lvov, V.~S. \& Falkovic, G.} 1992 {\em Kolmogorov Spectra
  of Turbulence I – Wave Turbulence\/}. Berlin: Springer.

\bibitem[Zhou(1995)]{Zhou95}
{\sc Zhou, Y.} 1995 A phenomenological treatment of rotating turbulence. {\em
  Phys.\ Fluids\/} {\bf 7}, 2092--2094.

\end{thebibliography}
\end{document}